\documentstyle[aps,prb,twocolumn,epsfig]{revtex}
\newcommand{\glsup}[1]{#1^{\raisebox{-.3ex}[.2ex][0ex]{$\scriptscriptstyle>$}
\atop   \raisebox{.3ex}[0ex][-.4ex]{$ \scriptscriptstyle<$} } }

\newcommand{\beq}{\begin{equation}}
\newcommand{\eeq}{\end{equation}}
\newcommand{\bea}{\begin{eqnarray}}
\newcommand{\eea}{\end{eqnarray}}

\renewcommand{\theequation}{\arabic{section}.\arabic{equation}}
\renewcommand{\appendix}{\renewcommand{\thesection}{Appendix \Alph{section}}%
\setcounter{equation}{0}%
\renewcommand{\theequation}{\Alph{section}\arabic{equation}}\section}

\draft
\begin{document}
\title{Nonequilibrium theory of scanning tunneling spectroscopy via
adsorbate resonances: nonmagnetic and Kondo impurities} 
\author{M. Plihal and J. W. Gadzuk}
\address{National Institute of Standards and Technology,
 Gaithersburg, MD 20899}

\maketitle
\thispagestyle{empty}
\begin{abstract}

We report on a fully nonequilibrium theory of the scanning tunneling
microscopy (STM) through resonances induced by impurity atoms adsorbed
on metal surfaces. The theory takes into account the effect of the
tunneling current and finite bias on the system and is valid for
arbitrary intra-adsorbate electron correlation strength. It is thus
applicable to the recent STM experiments on Kondo impurities. We
discuss the finite temperature effects and the consequences of
atomic scale resolution of the STM for the spectral property of such
systems. We find that the tip position affects the resonance
lineshapes in two ways. As a function of distance from surface, the
lineshapes vary due to the different extent of the adsorbate and metal
wavefunctions into the vacuum. However, we do not expect large
variations in lineshapes unless the tunneling into the tightly bound
adsorbate states is considerable or when nonequilibrium effects are
significant. As a function of lateral tip position, lineshapes should
not change significantly on length scales of $R_\parallel \le
10$~\AA$\,$ under typical experimental conditions when the electrons
tunnel into the perturbed bulk conduction states hybridized with the
outer shell $sp$ adsorbate orbitals. Tunneling into surface states on
(111) surfaces of noble metals should be important for observation of
the resonance at larger distances ($> 10$~\AA) and oscillatory
variations in the lineshape should develop. This long range behavior
has not been resolved in the recent experiments with Kondo
impurities. The temperature dependence of the Kondo resonance cannot
be deduced directly from the differential conductance as the thermal
broadening of the tip Fermi surface produces qualitatively similar
effects of comparable and larger magnitude. Careful deconvolution is
necessary to extract the temperate dependence of the Kondo
resonance. The finite-bias current-induced nonequilibrium effects in
tunneling through Kondo impurities should produce characteristic
broadening of the resonance in the case of strong hybridization of the
discrete state with the STM tip.

\end{abstract}
\pacs{PACS numbers:61.16.Ch,72.10.Fk}

\narrowtext

\section{Introduction}

A considerable body of experience and wisdom within the area of solid state
tunneling phenomenon was built up throughout tunneling's ``Golden Era of the
Sixties''.  It was during this period that many of the defining fundamental
ideas, basic theoretical strategies and methodologies, and broad scope of new
applications for tunnel structures were first realized.  A general introduction
to many of these achievements can be found in a number of
comprehensive volumes~\cite{Burstein69BOOK,Duke69BOOK,Wolf85BOOK} and
in the Nobel Lectures of Esaki, Giaver, and Josephson, who were
awarded the 1973 Nobel Prize in Physics for ``their [independent]
discoveries regarding tunneling phenomena in
solids''~\cite{Esakietal92BOOK}.  It is against this background that
the astounding achievements in contemporary tunneling studies
utilizing the single atom spatial resolution of the scanning tunneling
microscope (STM) are most meaningfully
considered~\cite{Young72RevSciIns,Binnig87RevModPh,Stroscio93BOOK,Wiesendanger93BOOK,RWi94,JulianChenBOOK}.   
One phenomenon of key interest here which was first considered in the
``Golden Era'' is that of impurity/adsorbate-assisted elastic
tunneling.  Two bodies of work are particularly relevant to the
present study.  The first is the recognition by Appelbaum and
coworkers of the possible role of the Kondo
effect~\cite{Kondo64ProgTheorPhys,Kondo69SSP,Hewson93BOOK} 
in determining certain current-voltage characteristics (e.g. ``zero-bias
anomalies'') of metal-oxide-metal tunnel junctions containing
localized paramagnetic impurity states near the metal-oxide
interfaces~\cite{Appelbaum67PR}. 
Second are the resonance tunneling studies involving valence electronic levels
of single atoms adsorbed on metal surfaces, as probed in a field emission 
microscope configured for energy analysis (thus enabling  electron
spectroscopy) of the field emitted
electrons~\cite{Duke67JCP,Plummeretal69SSC,GadzukPlummer73RevModPh}.
Many years later, useful parallels between the theory of single atom
resonance tunneling developed in the ``Golden Era'' and the theory of
the STM, in the single-atom-tip limit, were unambiguously
established~\cite{Gadzuk93PRB}.  Further discussion of these issues 
from the past will be offered throughout the text, when appropriate.

The basis for continuing interest and excitement in impurity/adsorbate-
assisted tunneling is that the transparency of tunnel
junctions can be dramatically enhanced by the presence of states localized
within the barrier when they are in resonance with the tunneling
electrons. Tunneling through such states is, for example, the origin of
conductance fluctuations quantum dots exhibit in the Coulomb 
blockade~\cite{GrabertDevoret92BOOK}. The tunneling probability in the
presence of a ``barrier'' state is proportional to the spectral
density produced by the hybridization of the localized state with the
conduction electrons and in many situations the current is given by
the Breit-Wigner formula 
\begin{equation}
I \propto \frac{\Gamma^2}{(\omega-\epsilon_0)^2+(\Gamma/2)^2}.
\end{equation}
Here $\Gamma$ is the width at half maximum of the resonance produced
by the hybridization with the conduction electrons in the right and
left lead and $\epsilon_0$ is the energy of the local state. The value
of $\Gamma$ depends on the height and width of the barrier potential between 
the central region and the leads.

Recently, enhancements in the zero bias conductance in quantum dots
due to the Kondo effect have been
observed~\cite{Goldhaber98Nature}. It had been shown earlier that 
$(dI/dV)$, the zero-bias differential conductance, is proportional to
the Fermi level density of states of the Kondo resonance on the
quantum dot. Similarly, the Kondo resonance has been spectroscopically 
observed on single magnetic impurities adsorbed on metal surfaces
using the STM~\cite{MadhavanetAl98S,LietAl98PRL}. However in the case
of the spectroscopic STM experiments, the resonance at the Fermi level
appears to have an asymmetric shape and cannot be interpreted simply
in terms of the local density of states of the impurity atom. Rather,
the electron tunneling current -- being a coherent quantum effect --
is a result of interference between competing tunneling channels, as
will soon be detailed.  Unlike in the quantum dot where the tunneling
can take place with appreciable magnitude only through the quantum dot
region, the apparent tunneling current from the STM tip to the surface
can either go through the resonance localized on the impurity or
directly into the conduction states of the surface. The distinction
between the conduction and local states will be discussed later. The
notion that the tunneling conductance is proportional to the local
density of states near the STM tip must then be modified.

In addition to its most common use for observing/determining atomic
geometrical structure at surfaces,
the STM is used as a sensitive probe
of surface electronic structure. Various theoretical approaches to the
STM conductance employ the tunneling Hamiltonian introduced by
Bardeen~\cite{Bardeen61PRL} and Golden Rule type expressions in which
under certain limiting conditions of practical interest the STM
conductance is indeed determined by the surface density of electronic
states near the STM
tip~\cite{TersoffHamman85PRB,GaoPerssonLundqvist97PRB}. Tersoff and
Hamann~\cite{TersoffHamman85PRB} developed a widely used model of the
scanning tunneling microscope that includes the three dimensionality
and spatial resolution of the tip.

The generic problem of a discrete state interacting with a continuum of states
arises in many different areas of physics and chemistry~\cite{MahanBOOK}. In
condensed matter
physics a frequently occurring realization is the electronic state of an
impurity atom immersed within a host
lattice.~\cite{KosterSlater54PR,Anderson61PR} In the case of magnetic
impurities, the interaction gives rise to nontrivial phenomena such
as the Kondo effect~\cite{Kondo64ProgTheorPhys,Kondo69SSP,Hewson93BOOK}. 

Within the context of atomic physics, Fano discussed related effects, 
as they might appear on observable absorption lineshapes or resonant
electron scattering cross sections which are due to the configuration
interaction (CI) that couples a discrete two-electron excited atomic
state with a continuum of ionization
states~\cite{Fano61PR,Cohen-Tannoudji92BOOK}. Whilst the ``natural'' 
lineshape of the resonance is Lorentzian, when studied  by
experiments in which an external probe interacts with the system, the
resonance can appear to have an asymmetric lineshape. Such lineshapes
are referred to as Fano resonances. Fano found that an asymmetry in
absorption lineshapes is due to interference between the excitation or
decay into CI-mixed discrete and continuum states which both couple to
the external probe. If the coupling between the probe and continuum is
expressed in terms of an energy independent matrix element $t_c$ and
the interaction between the probe and the localized state (which has
already been diluted by admixture into the continuum) by the matrix
element $\tilde t_a$, then the lineshape detected has the form 
\begin{equation}
\label{eq:Fano}
I \propto \frac{(q+2(\omega-\epsilon_0)/\Gamma)^2}{1+(2(\omega-\epsilon_0)
/\Gamma)^2},
\end{equation}
where $q\equiv \tilde t_a/(2\pi V t_c)$ with $V$ being the hybridization 
(or CI)
matrix element between the local state and continuum. The latter coupling
results in the
discrete state acquiring a width
$\Gamma = 2\pi \rho_s V^2$ ($\rho_s$ is the density
of continuum states).

In the present paper, we consider the problem of the discrete state
embedded in a continuum using a probe such as the STM that has atomic scale
spatial resolution. This work has been motivated by the recent STM 
experiments involving single Kondo
impurities~\cite{MadhavanetAl98S,LietAl98PRL}. The resonance
observed in the conductance was interpreted by the authors in
terms of the Fano interference. The fit of the resonance to the Fano
formula~\cite{MadhavanetAl98S} -- generalized to the case where 
intra-atomic Coulomb interaction on the impurity is taken into account
-- was based on the assumption that the tunneling into an Anderson
impurity can be extended to include the
tunneling into the continuum in a straightforward way. Upon further
consideration, it appears that this generalization of the Fano result to
the case of STM conductance is not as straightforward as has been
presumed. In the present paper we obtain a more complicated expression
than the elementary Fano formula, one which accounts for
the correct asymptotic behavior for large tip-impurity separation. In
particular, when the dependence of the probe's distance from the local
state is properly included, then observable consequences of the local state
admixture with the conduction electrons show the correct asymptotic
long range behavior in the large tip-to-impurity-separation-limit as
they obviously should.

The difference between the Fano lineshape and  the result obtained
here is due to the different nature of the probe in the ``multi-center" STM
configuration (one ``center'' on the impurity/discrete state, the other on
the STM tip/probe)
compared with the ``single-site'' atomic physics processes.  
While Fano was concerned with light absorption or electron scattering where 
the system under study was always at the probe's focus, the outcome of STM 
experiments must depend upon the variable spatial position of the
probe (tip) with respect to the discrete state under investigation.
Put another way, for the atomic physics applications considered by
Fano, both the discrete state coupled to the continuum and also the
initially excited decaying or autoionizing state (the ``probe'' state
in our STM language) are atomic states spatially localized at the same
site by the same atomic central potential. In contrast, the
``S''(=scanning!) in STM assures that the tip, hence initial excited
state, can be independently located with respect to the position of
the ``discrete state coupled to the continuum'' and it is this extra
degree of freedom that enriches the potential information content in
STM lineshape analysis, but also requires a much more detailed
theoretical treatment than merely fitting the atomic physics Fano
lineshape, Eq.~(\ref{eq:Fano}), to position-dependent STM
spectra. This will be expanded upon in depth later. This realization
demonstrates the importance and crucial need for considering the
measurement process in quantum mechanical observations. 

While the structuring of this comprehensive paper is based on a logical
development of the subject matter, it may be useful to present a roadmap of
key points and results to guide the °casualý as well as the °dedicatedý
reader. In section~\ref{sec:modelandapprox}, we introduce the model of
the system and discuss our approximations of the tunneling matrix
elements resulting in Eqs.~(\ref{eq:vdk}), (\ref{eq:tkp}), and
(\ref{eq:tdp2}). In section~\ref{sec:theorycurrent} and the supporting
appendices (A-C), we develop the general nonequilibrium theory of STM
tunneling current and conductance in the presence of an adsorbate
induced resonance. The more familiar equilibrium limit, asymptotically
exact for large tip-surface separation but pragmatically useful even
for moderate $\sim 5$\AA$\,$ separations, is treated as a special case 
in~\ref{subsec:tunnelcurrentequilibrium}. The relationship between the
equilibrium tunneling resonance lineshape observable in STM
experiments and the asymmetric Fano lineshape is established
in~\ref{subsec:tunnelcurrent} (Eqs. 3.24 and 3.25). The nonequilibrium
contribution to the lineshape treated as a correction to the
equilibrium limit is taken up in~\ref{subsec:nonequilibrium}. The
differential conductance is introduced
in~\ref{subsec:differential}. The crucial role of the substrate
electronic structure is explicitly considered in
section~\ref{subsec:bandaverage} where the substrate Green's function 
is evaluated for a jellium surface and
in~\ref{subsec:surfacebandaverage} where surface states and real
electronic structure effects are discussed qualitatively.

In section~\ref{sec:discussion}, we illustrate the predictions and
consequences of our theory on two models for the adsorbate:
nonmagnetic and Kondo. Using a jellium substrate model, we first
discuss the common features of the two adsorbate models in terms of
the non-interacting Anderson model in~\ref{subsec:nonint}. Four
different families of spectroscopic lineshape variations are taken up:
dependence on (1) the relative strength of the tip-to-adsorbate
vs. tip-to-substrate tunneling and on the substrate electronic
structure; (2) tip-surface separation; (3) lateral tip position; (4)
temperature due to Fermi level smearing. We explain why no variations
in the resonance lineshape should be observed for lateral tip
positions on length scales $\sim (1-2)$\AA$\,$ -- characteristic of
the bulk $k_F$ -- and only small lineshape variations should be
expected for vertical tip variations (in experimentally accessible
range). We leave the discussion of the tip and bias effects on
lineshapes to section~\ref{subsubsec:Kondononeqcurrent}.

Tunneling characteristics specific
to Kondo systems are presented in~\ref{subsec:Kondo}. We begin with
conceptual issues in ~\ref{subsubsec:conceptual}. We then discuss the
recent experiments on $Co/Au(111)$~\cite{MadhavanetAl98S} and
$Ce/Ag(111)$~\cite{LietAl98PRL} and relate our work to 
related theoretical
papers~\cite{Kawasakaetal99JAP,SchillerHershfield00PRB} 
(~\ref{subsubsec:Kondoequilibrium}). In particular, we show that the
temperature dependence of the Kondo resonance is not easily
extractable from the temperature dependence of the differential
conductance. We reiterate that the stability of the experimental
lineshape with the tip position is to be expected. Variations in the
lineshapes would, however, occur at larger distance due to tunneling
into the surface state. The effect of adsorbate-tip
hybridization and bias induced nonequilibrium on the current
(conductance) vs. bias measurements in Kondo systems are dealt with in 
section~\ref{subsubsec:Kondononeqcurrent}. Finally, an enumeration of
specific conclusions is offered in section~\ref{sec:conclusions}.

\section{Model and approximations}
\setcounter{equation}{0}
\label{sec:modelandapprox}

Models of scanning tunneling microscopy are abundant in the literature
of the last two decades and standard texts
exist~\cite{Stroscio93BOOK,Wiesendanger93BOOK,RWi94,JulianChenBOOK}.
We approach the problem as a nonequilibrium process and discuss the
corrections to the Tersoff-Hamann formulation~\cite{TersoffHamman85PRB}.
Our intent is to develop such theory under general and self-consistent
assumptions that accurately capture most of the qualitative aspects
involved and do so in a way that make extension to more realistic
calculations formally straightforward. We focus on the tunneling
through adsorbate resonances. Throughout this paper, we adopt the
convention that the energies are measured with respect to the
respective Fermi levels of the substrate and tip unless specified
otherwise and set $\hbar =1$. When the tip is biased we explicitly
shift the tip energies. 

\subsection{Model of the studied system}

We consider a system which consists of a clean
metallic surface with a single impurity atom adsorbed on it. The STM
will be used to study the system by means of tunneling through a
resonance produced by an electronic state of the impurity, such as the
$5f$ orbital of $Ce/Ag(111)$~\cite{LietAl98PRL} or $3d$ orbital in
$Co/Au(111)$~\cite{MadhavanetAl98S}. Unless otherwise noted, we place
the origin of the coordinate system at a point on the surface of the
metal directly below the adsorbate. This means that the position of
the impurity is $\vec R_0=(0,0,Z_0)$. The system without the probe is
described by the degenerate Anderson Hamiltonian
\begin{eqnarray}
\label{eq:SystemHamiltonian}
H_s(\vec R_0) &=& \sum_a \epsilon_0(\vec R_0) c^\dagger_a c_a +
\sum_{a > a'} U(\vec R_0) n_a n_{a'} + \\ \nonumber
&+& \sum_{ka} \epsilon_k c^\dagger_{ka} c_{ka} +
\sum_{ka} \left \{ V_{ka}(\vec R_0)
c^\dagger_{ka} c_a + {\mathrm H.c.} \right \}.
\end{eqnarray}
Here, $\epsilon_0$ is the energy of the impurity state $\psi_\sigma(\vec r)$,
which we assume may be a multiplet of states described collectively by
the quantum number $a\equiv(m\sigma)$. In the simplest case, $a$ correspond 
to the spin $\sigma$, but it may also include orbital degeneracy ($m$)
in more complicated cases. In this paper, we discuss at most spin
degenerate states with $a=\sigma$ and $N=2$ (degeneracy). We denote by
$c^\dagger_a$ the creation operator for this state. The $\epsilon_k$
is the conduction band state energy -- independent of $\sigma$ in the
absence of magnetic field -- with $c^\dagger_{ka}$ being the creation
operator for the corresponding Bloch state with symmetry (spin) $a$
common with the impurity state, and $V_{ka}$ is the matrix element for 
hybridization between the impurity and conduction states. The second
term in~(\ref{eq:SystemHamiltonian}) corresponds to the intra-atomic 
Coulomb interaction between electrons in the impurity state
$\psi_a$.

\begin{figure}
\centerline{\epsfxsize=0.4\textwidth
\epsfbox{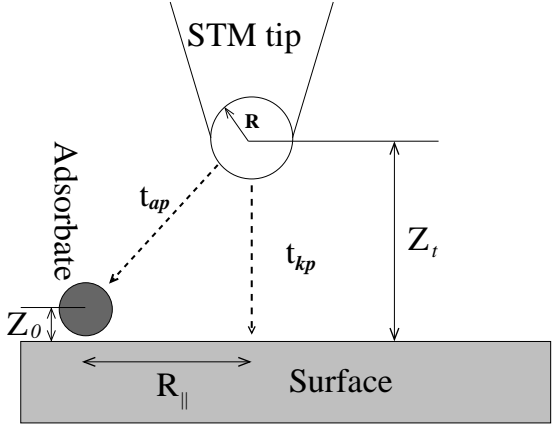}
}
\caption{ schematic picture of the STM }
\label{fig:1}
\end{figure}
If the renormalized energy $\epsilon_0$ lies within the conduction
band, the bound state broadens into a resonance which in the wide-band limit
and with $U=0$, has a Lorentzian shape
\begin{equation}
\label{eq:Lorenzian}
\rho_0(\omega)=\frac{1}{2\pi}\frac{\Gamma}{(\omega-\epsilon_0)^2+(\Gamma/2)^2}, 
\end{equation}
where
\begin{equation} \label{eq:widthwideband}
\Gamma = 2\pi \rho_s V^2,
\end{equation}
with $\rho_s$ and $V$ the assumed-energy-independent density of conduction
states and hybridization matrix element
from~(\ref{eq:SystemHamiltonian}). In most of our later considerations
however, we will retain an energy dependence in the model density of
states, $\rho_s(\omega)$.

\subsection{Interaction of the system with the STM tip}

When the STM tip is brought near the impurity, electrons can tunnel
between the tip and the adsorbate state. This situation is expressed
by adding an interaction term to the Hamiltonian
\begin{equation}
\label{eq:tunnelingd}
H_{at}(\vec R_t,\vec R_0)=\sum_{pa} \left \{ t_{ap}(\vec R_t,\vec R_0) 
c^\dagger_a c_p + H.c. \right \}. 
\end{equation}
The tip states are
denoted by subscript $p$ while the conduction states of the metal by
subscript $k$. The transfer matrix $t_{ap}$ depends on the
position of both the adsorbate $Z_0$ and the tip $\vec R_t \equiv
(\vec R_\parallel, Z_t)$. 

If the STM only coupled to the discrete state with transfer amplitude
$t_a$ then the conductance would, in the wide-band
limit ($t_{ap}\equiv t_a$, independent of $p$), be determined by
\begin{equation}
\label{eq:conductancetoresonanceonly}
G \propto |t_a|^2\frac{\Gamma}{(\omega-\epsilon_0)^2+(\Gamma/2)^2}
\end{equation}
and the conductance would thus be directly related to the impurity
density of states. This is reminiscent of the defining characteristics
from field emission resonance tunneling spectroscopy in which tunneling
from the substrate to vacuum is disproportionately smaller than that from
``good''
adsorbates~\cite{Duke67JCP,Plummeretal69SSC,GadzukPlummer73RevModPh,Pennetal72PRB}. 
However, since in the STM geometry, tunneling directly between the tip
and the metal surface can be comparable to (or in excess of) that
between the tip and the impurity, the conductance exhibits a more
complex behavior than that of a simple impurity local density of
states. We take such processes into account through 
\begin{equation}
\label{eq:tunnelingk}
H_{st}(\vec R_t)=\sum_{pk} \left \{ t_{kp}(\vec R_t) c^\dagger_k
c_p + H.c. \right \},
\end{equation}
where the tunneling matrix element $t_{kp}$ depends on the position of
the tip $\vec R_t$. In the rest of this section, we will address the
issue of the probe's effect on the system itself. This is particularly
important when the tip is brought very close to the adsorbate so that
the tunneling $t_{ap}$ and $t_{kp}$ are comparable with $V_{ka}$. In
this case, the width $\Gamma$ of the adsorbate resonance is no longer
given by~(\ref{eq:widthwideband}) but 
rather by
\begin{equation}
\Gamma=2\pi (\rho_s V^2 + \rho_t t_{a}^2 )
\end{equation}
The effect of the tip on the conduction states can also be important.
However, the most important consequence of the tip perturbation comes
when finite bias is applied across the tunnel junction. The system
will be out of equilibrium and the problem of the system-probe must be
approached self-consistently with the tip included in the system it is
probing. The typical operational mode of the STM during imaging and
spectroscopic measurements is such that the tip distance from the
adsorbate is several atomic units larger than the adsorbate-surface
separation and therefore $|t_{ap}|, |t_{kp}| \ll |V_{ka}|$. In this limiting case,
the probe has no effect on the adsorbate-metal complex other than as a 
source of hot tunneling electrons. This is a reasonable assumption as long
as the tip separation is large enough to justify the approximation
$|t_{ap}|, |t_{kp}| \ll |V_{ka}|$.

\subsection{Model of the STM tip}

An important property of the tip is its spatial
resolution, as discussed in great detail by 
many~\cite{Binnig87RevModPh,Gadzuk93PRB,TersoffHamman85PRB,Chen90PRL,Chen91JVacSci,Langetal89PRL}.
We will consider the tip to be well defined and terminated by a single
atom through which the tunneling predominantly takes place.  This is
the s-wave model of Tersoff and Hamann~\cite{TersoffHamman85PRB}, see
Fig~\ref{fig:1}. The important features are the following: (a) the tip
Hamiltonian is 
\begin{equation}
\label{eq:tipHamiltoniandiagonalized}
H_t = \sum_{p} \epsilon_p c^\dagger_p c_p,
\end{equation}
where $c^\dagger_p$ creates an electron in the state $\psi_p(\vec r)$
with energy $\epsilon_p$ measured from the Fermi level of the tip $\epsilon_{Ft}$;
(b) when the tip is positioned near the surface, tunneling into and
out of a state $\psi_p$ can take place; (c) the states are filled up to
the chemical potential $\epsilon_{Ft}$ controlled by the bias; (d)
the tip states are characterized by a density of states which we
denote by $\rho_t(\omega)=\sum_p \delta (\omega -
\epsilon_p)$; and (e) the asymptotic form of the tip
eigenstate $\psi_p$ in the vacuum region extending towards the metal
surface is characterized by the atomic orbital of the apex atom. The
wavefunction $\psi_p$ can be found based on simple physical arguments without 
solving the complete problem. If $\phi_t$ is the work function of the
tip and $\kappa_t\equiv \sqrt{2m^*_t (\phi_t-\epsilon_p)}$, then
following Tersoff and Hamann, 
\begin{equation}
\label{eq:tipwf}
\psi_p(\vec r) \propto R \, e^{\kappa_t R} \;
\frac{\exp(-\kappa_t r)}{r}
\end{equation}
where R is the radius of curvature of the tip about its center which
is located at the origin of this ``tip-defining'' coordinate system.
While ~(\ref{eq:tipwf}) represents an ``s-wave tip'', more generally
$\psi_p$ would carry
whatever symmetry was possessed by the relevant atomic orbital centered
at the tip apex~\cite{JulianChenBOOK,Gadzuk93PRB,Chen90PRL}. The
wavefunction tail, controlled by $\kappa$, depends
on the bias and tip-surface separation. Both factors modify the height
of the vacuum barrier, hence effective work function $\phi$ determining
$\kappa$. These modification can be essentially included by
renormalizing the wavefunction tails and the densities of states by
position and energy dependent factors via the tunneling matrix
elements. We discuss the tunneling matrix elements next. 

\subsection{Approximations for the tunneling matrix elements}
\label{subsec:theorytunneling}

An important role in our formulation is played by the tunneling
(hybridization) matrix elements $V_{ka}$, $t_{ap}$, and $t_{kp}$ since
they include the dependence on electronic structure and the tip and
adsorbate position. The desired quantitative accuracy of the model for the
tunneling process is to a large degree determined by the
approximations made in the evaluation of these matrix
elements.~\cite{Gadzuk93PRB,Chen90PRL,Wille94PRB} We do this first
for a general adsorbate-metal system and then for a jellium model. It
is rather straightforward to include band structure effects using a
realistic electronic structure calculation of the substrate Green's
function.

We begin with the discussion of the matrix elements $V_{ka}$ and
$t_{kp}$ that contain the metal wavefunctions $\psi_{\vec k}(\vec
r)$. They have the form
\begin{equation}
\label{eq:Vdk}
M_{kl}(\vec R_l)= \int d^3r \psi^*_{\vec k}(\vec r)
v_{sl}(\vec r; \vec R) \psi'_l(\vec r),
\end{equation}
where $v_{sl}$ is the potential representing the mutual interaction 
of the two systems. The wavefunction $\psi_{\vec k}(\vec r)$ is a
Bloch state of the unperturbed metal and $\psi'_l (\vec r)$ is the
wavefunction (in the coordinate system of the metal) of the
adsorbate state $a$ in the case of $V_{ka}$ or the tip wavefunction in
the case of $t_{kp}$. Either way, $\psi'_l (\vec r)$ can be written in
terms of the wavefunction with the origin at the adsorbate (tip) as
$\psi'_l(\vec r)=\psi_l(\vec r-\vec R_l)$, where $\vec R_l$ is the
position of the adsorbate ($\vec R_0$) or the tip apex atom ($\vec
R_t$) measured from a reference point on the surface. The energies
$\epsilon_k$ of the metal electrons are written in terms of the
perpendicular and parallel components as $\epsilon_k \equiv
\epsilon_{k_z}+\epsilon_{k_\parallel}$. We follow the convention that
$\epsilon_{k_\parallel}$ is measured from the bottom of the 2-D band
and $\epsilon_{k_z}$ is measured with respect to $\epsilon_{Fs}$. For
example, we write for the jellium model
$\epsilon_{kz}=k_z^2/2m^*_s-D$ and
$\epsilon_{k_\parallel}=k_\parallel^2/2m^*_s$, where $(-D)$ is the
energy of the bottom of the band with respect to the Fermi level. The
Bloch states can generally be written in the relevant region outside
the metal as 
\begin{equation} 
\label{eq:metalstate}
\psi_{n\vec k}(\vec r)=
e^{-\kappa_{ns} z} u_{n\vec k_\parallel}(\vec \rho,z)
e^{i\vec k_\parallel \cdot \vec \rho}.
\end{equation}
where $n$ is the band index, $u_{n\vec k_\parallel}(\vec \rho,z)$ is
a function weakly dependent on $z$ outside the surface and periodic
in $\vec \rho$, the electron coordinate in the plane of the surface.
This form is equally valid for the metal band gap surface states that
seem ubiquitous to STM studies on (111) noble metal
surfaces since both the z-propagating Bloch states and the localized
surface states are eigenstates of the same Hamiltonian with different
eigenvalues at a given $k$. At the surface, evanescent states into the
bulk that appear upon analytic continuation of the band structure into
the domain of complex k-vectors cannot be rejected on the basis of
physical considerations as they were for the perfectly periodic
interior of the solid. 

For jellium, $u_{n\vec k_\parallel}(\vec r)$ is constant and the 
the metal states are then simply plane waves along the surface with
exponentially decaying amplitude into the vacuum. Here $\kappa_{ns} =
\sqrt{2m_s^* (\phi_s - \epsilon_{nk_z})}=\sqrt{2m^*_s (\phi_s -
\epsilon_{nk} + \epsilon_{nk_\parallel})}$ with $m^*$ the metal
electron effective mass (number) and $\phi_s$ the height of the
tunneling barrier for a Fermi level electron. For bias voltages much
smaller than the work function, $\phi_s$ is equal to the metal work
function. We omit the band index $n$ in the rest of the paper unless
we explicitly discuss the electronic structure effects. We also define
$\lambda_\omega^{-1} =\sqrt{2m_s^*(\phi_s-\omega)}$, where the energy
factor $(\phi_s-\omega)$ represents the effective tunneling potential
barrier for an electron with energy $\omega$. For small bias voltages,
$\omega \approx 0$, and we can replace $\lambda_\omega$,
which depends weakly on energy in this range, by its Fermi level value 
$(\equiv \lambda)$. 

We shift the integration variable in the integral~(\ref{eq:Vdk}) to an
origin centered on the adatom (tip), which gives 
\begin{eqnarray}
\label{eq:Vdkapproximate}
M_{kl}(\vec R_l)&=&
\psi^*_k(\vec R_l) 
\int_{z>-Z_l} d^3r e^{-\kappa_s z}e^{-i \vec k_\parallel \cdot \vec
\rho} \times \\ \nonumber
&\times& \frac{u^*_{\vec k}(\vec r+\vec R_l)}{u^*_{\vec k}(\vec R_l)}
v_{sa}'(\vec r,R_l) \psi_l(\vec r).
\end{eqnarray}
Matrix elements of this type have been the focus of intense study in
the context of charge transfer processes at
surfaces.~\cite{Wille94PRB,Gadzuk67SS,NordlanderTully90PRB,Borisovetal96PRA,Sautet97SS,Plihal99PRB1}
The main contribution to the integral
in~(\ref{eq:Vdkapproximate}) comes from the region just outside the
surface where the $z$-dependence of the metal states is that of a
decaying exponential. In the case of both the adatom and the tip, the
integrand is assumed to be well localized to the $z>-Z_0$ region by
virtue of the spatial properties of $\psi_l$. Consequently, the
integral will be reasonably constant for all $k_\parallel \leq 1/r_l$
where $r_l$, the radial length scale for the atom or tip, sets the
range of $k_\parallel$. Since, as will soon be demonstrated, other
factors in the full problem will provide a much more severe
$k_\parallel$ cutoff, it is sufficient to represent the integral by a
constant. We then write the matrix element approximately $M_{kl}(\vec
R_l) \simeq M_0 \psi^*_{\vec k}(\vec R_l)$, where $M_0$ is the overlap
integral defined in~(\ref{eq:Vdkapproximate}) that contains the
dependence of the matrix element on the symmetry of the atomic
orbitals near the tip and on selection rules.

Since the distance of the adsorbate from the surface is small, it will
be sufficient for the purpose of $V_{ka}$ to write the Bloch state in
the form $\psi_{\vec k}(Z_0)=e^{-Z_0/\lambda} \psi_{\vec k}(0)$ with
the decay constant independent of the k-vector. With the choice of our
coordinate system $\vec R_0=(0,Z_0)$, we write $V_{ka}$ in the form of
a separable product of a k-independent function of adsorbate position
multiplied by a function of $\vec k$
\begin{equation}
\label{eq:vdk}
V_{ka}(\vec R_0) \simeq V_a(Z_0) \psi^*_{\vec k}(0) 
\end{equation}
where $V_a(Z_0)=V_0 e^{-Z_0/\lambda}$. We cannot make this last
approximation in the tip-to-surface tunneling matrix element $t_{kp}$,
since the interplay between $k$ and $Z_t$ dependence has an
important role in the tunneling process. However, from the conceptual
point of view, we find it convenient to isolate a $k$-independent
dependence on the tip position in $t_c(Z_t)=t_0 e^{-Z_t/\lambda}$ and
write the matrix element in the form
\begin{equation}
\label{eq:tkp}
t_{kp}(\vec R_t)=t_c(Z_t) e^{Z_t/\lambda} \psi^*_{\vec k}{(\vec R_t)}.
\end{equation}
We note that the s-wave tip-to-surface tunneling matrix element $t_{kp}$ has been
shown by Tersoff and Hamann~\cite{TersoffHamman85PRB} under quite
general assumptions to assume the form~(\ref{eq:tkp}) with the tip
wavefunction given by Eq.~(\ref{eq:tipwf}), which indicates that our
simple qualitative arguments seem to be supported by more detailed
analysis. 

The tip-to-impurity matrix element 
\begin{eqnarray}
\label{eq:tdp}
&&t_{ap}(\vec R_t, \vec R_0)= \\ \nonumber
&& = \int d^3r \; \psi^*_{\vec p}(\vec r - \vec
R_t) v_{at}(\vec r;\vec R_t,\vec R_0) \psi_a(\vec r-\vec R_0). 
\end{eqnarray}
depends on the position of both the adsorbate ($Z_0$) and the tip
($\vec R_t$). Since the tunneling from the tip takes place
predominantly through the apex atom, the wavefunction
$\psi_p$ in the last expression, a generalization of
~(\ref{eq:tipwf}), can be written (with the tip at origin) as 
\begin{equation}
\label{eq:psitip}
\psi_{\vec p}(\vec r) =e^{\kappa_t R} \; 
\frac{e^{- \kappa_t r}}{r} \psi_s(\hat r)
\end{equation}
where $\psi_s$ is the angular part
of the orbital localized on the apex atom whose center is located at
$\vec R_t$. The tunneling matrix element $t_{ap}$
will thus reflect the symmetry and spatial dependence of the states 
localized on the adsorbate and in the tip apex. Within the s-wave tip
model, $\psi_s$ is just an innocuous constant.  Since $\psi_p$ and
$\psi_a$ appearing in Eq.~(\ref{eq:tdp}) are both atomic-like
functions in the relevant region of overlap centered respectively 
on the tip and on the adatom, in a broad sense $t_{ap}$ is similar to a
common two-center hybridization/hopping integral defining the binding
in a diatomic molecule~\cite{Eyringetal44BOOK,LevineBOOK}.
Typically the magnitudes of these integrals are exponentially
decreasing functions of their separation (possibly multiplied by a
mildly oscillatory function accounting for the nodal structure of the
atomic functions).  Based on this analogy from quantum chemistry,
$t_{ap}$ given by Eq.~(\ref{eq:tdp}) should take the form 
\begin{equation}
\label{eq:tdp2}
t_{ap}(\vec R_t, Z_0)\approx t_a e^{-|\vec R_t - Z_0 \hat i_z|/\alpha }
\equiv t_a(\vec R_t, Z_0)
\end{equation}
Here, $\alpha^{-1}\approx\left (\kappa_t +
\sqrt{2m(\phi_s-\epsilon_0)}\right )$ is an effective decay constant 
evaluated for states at the Fermi level of the tip, $|\vec R_t - Z_0
\hat i_z|= \sqrt{R^2_\parallel +(Z_t-Z_0)^2}$ is the tip-to-atom
separation, as depicted in Fig. 1, where $\vec R_\parallel$ is the parallel
component of $\vec R_t$. The decay constant $\kappa_t$ depends on the
energy of the tip state $(\epsilon_p)$, but this dependence is
very weak for small biases considered here and we neglect it. In this
case, the matrix element is well approximated by the $p$ independent
form $t_a(\vec R_t, Z_0)$. The matrix element $t_a$ may be taken
real. 

\section{The nonequilibrium theory of the tunneling current and
differential conductance} 
\setcounter{equation}{0}
\label{sec:theorycurrent}

The tunneling between the adsorbate-metal complex and a biased tip is
a nonequilibrium process. Although we frequently make the assumption
in this paper that the tip-system interaction is weak enough so that
local equilibrium is maintained to a good approximation, the
assumption is less valid when the tip is near the surface. For this
reason, we develop our theory of the tunneling process within the
Keldysh-Kadanoff~\cite{Keldysh65JETP,KadBaym} framework for the
nonequilibrium Green's functions and discuss the nonequilibrium
corrections. 

\subsection{General expression for the tunneling current in terms of
nonequilibrium Green's functions}
\label{subsec:generaltunneling}

We define the tunneling current as the flow of electrons through a
closed surface around the tip. It is expressed in terms of the
continuity equation as
\begin{equation}
\label{eq:continuityeq}
I = -e \langle \frac{d n_t(t)}{dt} \rangle
\end{equation}
where $n_t=\sum_pc^\dagger_p c_p$ is the number operator for
the tip electrons, and the brackets signify the ensemble average,
which in the local equilibrium case is the thermal average over the
tip states. The time derivative is found from the Schr\"odinger
equation of the total Hamiltonian of the tip-substrate-adsorbate
system $H_{tot} = H_s + H_t + H_{at} + H_{st}$. Since the number
operator commutes with $H_s$ and $H_t$, the only contribution comes
from the interaction terms $H_{at}$ and $H_{st}$ and the current is  
\begin{equation}
\label{eq:current}
I=\frac{2e}{\hbar} \;{\mathrm Im} \; \{ 
\sum_{k p} t_{kp} \langle c_k^\dagger(t) c_p(t) \rangle + 
\sum_{a p} t_{ap} \langle c_a^\dagger(t) c_p(t) \rangle
\},
\end{equation}
where we omitted the arguments in $t_{kp}$ and $t_{ap}$ for
convenience and used the relation $t_{ap}=t^*_{pa}$ and
$t_{kp}=t^*_{pk}$. We write the arguments explicitly only when we
wish to emphasize their dependence. We define the time loop Green's
functions $G_{pa}(t,t')=-i\langle T_C c_p(t) c_a^\dagger(t') \rangle$
and $G_{pk}(t,t')=-i\langle T_C c_p(t) c_k^\dagger(t') \rangle$, where
$T_C$ orders the times along a contour $C$ in the complex time
plane. The contour can be taken to be the Kadanoff-Baym
contour~\cite{KadBaym}, the Keldysh contour~\cite{Keldysh65JETP}, or a
more general choice. The discussion of nonequilibrium Green's function
is available in standard books and review
articles~\cite{KadBaym,LangrethBOOK,JauhoBOOK} and we refer the reader 
to these references for further details. Our notation follows closely
that of the more detailed discussion in
reference~\cite{Plihal98PRB}. In the present paper, we are mainly
interested in steady state tunneling current (time independent) and
therefore we work with the Fourier transformed quantities in frequency
rather than time space. The current can be written as
\begin{equation}
\label{eq:currentfrequency}
I=\frac{2e}{h} \;{\mathrm Im}
\int_{-\infty}^\infty d\omega \{ 
\sum_{k p} t_{kp} G_{pk}^<(\omega) + \sum_{a p} t_{ap}
G_{pa}^<(\omega) \},
\end{equation}
where $G^<_{pk}(\omega)$ and $G^<_{pa}(\omega)$ are the Fourier
transforms of $G_{pk}^<(t,t')=\langle c_k^\dagger(t') c_p(t) \rangle $ and 
$G_{pa}^<(t,t')=\langle c_a^\dagger(t') c_p(t) \rangle$ -- the
analytic pieces on the real time axis of the Green's functions
introduced above. Equivalently, the current may be calculated from
$\langle \frac{dn_s}{dt} + \frac{dn_a}{dt} \rangle$. It is easy to see
that this approach also leads to the equation~(\ref{eq:current}).

The problem of finding the current thus reduces to finding the
``lesser'' Green's functions $G^<_{pa}$ and $G^<_{pk}$. This is done
using the equation of motion method for the time ordered Green's
functions in \ref{app:equationofmotion} and the rules for analytic
continuation described in~\ref{app:analytic}. In order to see the
interference between the two scattering channels giving rise to the
Fano lineshape, it is also useful to express the current
using~(\ref{eq:greenpd}) and~(\ref{eq:greenpk}) in
Eq.~(\ref{eq:currentfrequency}) as  
\begin{eqnarray}
\label{eq:currentfirststep}
I&&=\frac{2e}{h}\;{\mathrm Im}\sum_p \int_{-\infty}^\infty d\omega
\\ \nonumber 
&& \{ G^{(0)}_p [ \sum_{a}|t_{pa}|^2
\;G_a + \sum_{ka} t_{ap} t_{pk} \;G_{ka} +
\\ \nonumber 
&&+ \sum_{ka} t_{kp} t_{pa} \;G_{ak}
+\sum_{kk'} t_{pk'} t_{kp} \;G_{k'k}] \}^<. 
\end{eqnarray}
The first term corresponds to the tunneling into the adsorbate state
hybridized with the metal electrons. The fourth term is the
contribution to the current from the direct tunneling into the
conduction states perturbed by the presence of the discrete adsorbate
state. The second and third terms give the interference between the two
channels.

We substitute the solutions~(\ref{eq:greenpdsimplified}) 
and~(\ref{eq:greenpksimplified}) for $G_{pk}$ and $G_{pa}$
from~\ref{app:equationofmotion} into
Eq.~(\ref{eq:currentfrequency}) and write the tunneling current in the form
\begin{eqnarray}
\label{eq:currentfinal}
I=&&\frac{2e}{h} \; {\mathrm Im} \int_{-\infty}^\infty
d\omega \times \\ \nonumber &\times& \sum_{pp'} \left \{ 
G_{pp'}^R(\omega) T_{p'p}^<(\omega) + 
G_{pp'}^<(\omega) T_{p'p}^A(\omega) 
\right \},
\end{eqnarray}
where the retarded function $T_{pp'}^R(\vec R_t,\vec R_0,\omega)$
plays the role of the T-matrix for scattering of the tip electrons
from the adsorbate-metal complex. It is defined by 
\begin{eqnarray}
\label{eq:Tpp'}
T_{pp'}&=& \sum_k t_{pk} G_k^0 t_{kp'} + \\ \nonumber
&+&\sum_a \tilde t_{pa} G_a \; (
t_{ap'} + \sum_k \tilde V_{ak} G_k^0 t_{kp'} ),
\end{eqnarray}
with 
\begin{equation}
\tilde V_{ka} = V_{ka} + \sum_p t_{kp} G_p^0 t_{pa}
\end{equation}
and 
\begin{equation}
\tilde t_{pa} = t_{pa} + \sum_k t_{pk} G_k^0 V_{ka}
\end{equation}
the hybridization and
tunneling matrix elements for the adsorbate modified by the tip-substrate
interaction. The matrix $T_{pp'}$ incorporates the properties of the tip as
well as the adsorbate into the expression for current. We discuss its physical
meaning more in the next section.

\subsection{Equilibrium limit of the tunneling current at large
tip-surface separation}
\label{subsec:tunnelcurrentequilibrium}

We define the equilibrium tunneling current as the large tip-surface
separation limit of~(\ref{eq:currentfinal}) when the tip, adsorbate, and
substrate are all in local equilibrium. This is equivalent to keeping
only the lowest order terms in $t_{kp}$ and $t_{pa}$. In our
formalism, this is achieved by replacing $\tilde G_{pp'}\rightarrow
G_p^0 \delta_{pp'}$ in Eq.~(\ref{eq:currentfinal}), $\tilde
V_{ka}\rightarrow V_{ka}$ in $T_{pp'}$, and by using the
fluctuation-dissipation relation $G_i^<(\omega)=f_i(\omega)
\rho_i(\omega)$. The subscript $i$ stands for tip (t), adsorbate (a),
and metal (s), respectively. The adsorbate is in equilibrium with the
metal, i.e. $f_a(\omega)=f_s(\omega)$. The matrix $T_{pp'}$ is
expressed entirely in terms of the Green's functions of the system and
the tunneling matrix elements $t_{pk}$, $t_{pa}$. Since these matrix
elements reflect the symmetry of the apex atom wavefunction, but are
only weakly dependent on $p$ on the energy scale of the resonance width,
the matrix $T_{pp'}$ will also have this property. We therefore make an
additional assumption that $\sum_p G_p^0 T_{pp} \sim (\sum_p G_p^0)
T_t$, where
\begin{eqnarray}
\label{eq:Tt}
T_t= \sum_k t_{pk} G_k^0 t_{kp} + \sum_a \tilde t_{pa} G_a \tilde t_{ap}
\end{eqnarray}
is only a function of the atomic tip orbital independent of
$p$. We define a tip-specific quantity observable by the STM, which is
related to the local density of states 
\begin{equation}
\label{eq:tiplocaldensity}
\tilde \rho_{sat}(\vec R_t, \vec R_0;\omega)=
- \frac{1}{\pi}{\mathrm Im}\,T_t^R(\vec R_t, \vec R_0;\omega),
\end{equation}
and write the equilibrium current $I_{eq}(\vec R_t, \vec R_0, V)$ as
\begin{equation}
\label{eq:currentfinalequilibrium}
I_{eq}=\frac{2e}{h} \int_{-\infty}^\infty
d\omega \left [ f_t(\omega') - f_s(\omega)  \right ]
\rho_t(\omega') \tilde \rho_{sat}(\omega),
\end{equation}
where $\rho_t= \sum_p \delta(\omega-\epsilon_p) $ is the
density of tip states and $\omega'=\omega-eV$ with $V$ being the bias
voltage. This equations is easily related to traditional formulations
given in terms of an integral product of an electron ``supply
function'' multiplied by a tunneling or transmission
probability~\cite{Burstein69BOOK,Duke69BOOK,Wolf85BOOK,Duke67JCP,GadzukPlummer73RevModPh}
when it is realized that $\tilde \rho_{sat}$, as defined here, already
contains within it factors ($\propto |t|^2$) representing the role of
the tunneling probability. 

This expression has a form similar to the standard tunneling theories 
which express the current as a product of the local densities of
states of the two systems evaluated at a common point and a difference
in the corresponding Fermi 
functions. Kawasaka et al.~\cite{Kawasakaetal98PLA,Kawasakaetal99JAP} 
who studied the STM current through a Kondo resonance used as a
starting point of their considerations the Tersoff and Hamann 
expression~\cite{TersoffHamman85PRB}
\begin{equation}
\label{eq:currentTersoffHamman}
I_{eq} \propto \int d\omega 
\left [ f_t(\omega') - f_s(\omega)  \right ]
\rho_t(\omega') \rho_{sa}(\vec R_t,\omega)
\end{equation}
according to which the current at zero temperature is
related to the LDOS of the adsorbate plus metal electrons
$\rho_{sa}(\vec R_t,\omega)$ at the position of the tip, where the
local density of states is
\begin{equation} 
\label{eq:localdensityofstates}
\rho_{sa}(\vec R_t, \omega) = - \frac{1}{\pi}\, {\mathrm Im} \;
\langle \vec R_t \, | G^R(\omega)|\, \vec R_t \rangle.
\end{equation}
The LDOS is expressed in terms of unperturbed
metal and adsorbate states by inserting $\sum_k | k \rangle \langle k
| + |a\rangle\langle a| (\approx 1)$ on both sides of $G$ 
in~(\ref{eq:localdensityofstates}). This is strictly valid only for
orthogonal orbitals, $\langle a | k \rangle=0$. The four resulting
terms, proportional to $G_a$, $G_{ka}$, $G_{ak}$, $G_{kk'}$, reflect
the fact that the LDOS includes both the adsorbate and metal
states perturbed by their mutual interaction. Inserting this expansion
into Eq.~(\ref{eq:localdensityofstates}) gives for
$\rho_{sa}(\vec R_t,\omega)$
\begin{eqnarray}
\label{eq:localdensity}
\rho_{sa}&=&-\frac{1}{\pi}{\mathrm Im} \;
\{ \sum_{a}|\psi_a'|^2 \;G_a^R +
\sum_{ka} \psi_{\vec k} \;G^R_{ka}\psi_a'^* + \\ \nonumber
&+& \sum_{ka} \psi_a'(\vec R_t)G^R_{ak}\psi^*_{\vec k}\ +
\sum_{kk'} \psi_{\vec k'} G^R_{k'k} \psi^*_{\vec k} \},
\end{eqnarray}
where the wavefunctions are evaluated at $\vec R_t$.
The four terms correspond to the terms in the square brackets
in~(\ref{eq:currentfirststep}). We can also rewrite $\rho_{sa}$ using 
the expressions for $G_{ak}$ and $G_{kk'}$ in~\ref{app:equationofmotion}.
We define $\tilde \psi_a = \psi_a' + \sum_k \psi_k G_k^0 V_{ka}$ and write 
\begin{equation}
\label{eq:localdensityfinal}
\rho_{sa}=-\frac{1}{\pi}{\mathrm Im} \;
\{ \sum_k|\psi_k|^2 \;G_k^R +
\sum_{a} | \tilde \psi_a|^2 \;G^R_{ka}
\}.
\end{equation}

Comparison of (\ref{eq:localdensityfinal}, \ref{eq:currentTersoffHamman})
with (~\ref{eq:Tt}, ~\ref{eq:currentfinalequilibrium}) shows the
difference between our equilibrium limit and the transfer Hamiltonian
method~\cite{Duke69BOOK,Bardeen61PRL}. The tunneling current and
differential conductance in the equilibrium limit provides information
about $\tilde \rho_{sat}$ -- a local density of states modified by the
tunneling matrix elements -- rather than the LDOS. In the case when
the tunneling takes place into distinct orbitals with different
symmetry, $\tilde \rho_{sat}$ can be rather different from $\rho_{sa}$
and the statement that the STM is a measure of local density of states
must be understood in this context. 

\subsection{The equilibrium tunneling current and the Fano lineshape}
\label{subsec:tunnelcurrent}

In this section, we evaluate $\tilde \rho_{sat}$ in
Eq.~(\ref{eq:currentfinalequilibrium}) and write the equilibrium
current using the approximations~(\ref{eq:vdk}), (\ref{eq:tkp}), and
(\ref{eq:tdp2}) for the tunneling matrix elements. We note that the
approximations do not require any specification of the substrate
electronic structure. The final form of the current allows the
discussion of the tunneling resonances in terms of the well
established Fano lineshapes. 

We introduce new quantities in terms of which the current is
expressed. First, we define the ``bulk'' density-of-states (DOS) for
the substrate and the tip as
$\rho_s(\omega)=\sum_k\delta(\omega-\epsilon_k)$ 
and $\rho_t(\omega)=\sum_p\delta(\omega-\epsilon_p)$. The impurity
width without the STM tip is defined as 
\begin{equation}
\label{eq:gamma1}
\Gamma_{as}(\vec R_0,\omega) = 2\pi \rho_s(\omega) V_a^2(\vec R_0).
\end{equation}
The adsorbate perturbation on the local density of conduction states 
at some lateral position between the tip and the adsorbate is
discussed in terms of the unperturbed substrate Green's function  
\begin{equation}
\label{eq:freemetalgreen}
G^+_0(\vec r, \vec r\,';\omega) = \sum_k
\frac{\psi_k(\vec r)\psi^*_k(\vec r\,')}{\omega-\epsilon_k+i \eta }.
\end{equation}
We define two dimensionless quantities related to the real and
imaginary parts of the Green's function
\begin{equation}
\label{eq:Lambda}
\Lambda(\vec R, \omega) = e^{Z/\lambda}
\frac{{\mathrm Re}\; G^+_0(\vec R, 0;\omega)} {\pi \rho_s(\omega)}
\end{equation}
and 
\begin{equation}
\label{eq:gamma}
\gamma(\vec R,\omega) =
-e^{Z/\lambda}\frac{{\mathrm Im}\; G^+_0(\vec R, 0;\omega)}
{\pi \rho_s(\omega)}.
\end{equation}
These two functions carry the information about both the spatial
extent of the metal electron perturbation at arbitrary $\vec R$ in the
surface region due to a localized perturbation at $\vec R_\parallel
=0$ and also the spatial resolution of the tip, as we will see
later. We have included the exponential factor $e^{Z/\lambda}$ in the 
definition~(\ref{eq:Lambda}),~(\ref{eq:gamma}),
and~(\ref{eq:nudef}) because we explicitly take the $k$-independent
part of the exponential dependence on position to be part of the
tunneling matrix elements $V_a(Z_0)$, $t_a(\vec R_t, Z_0)$, and
$t_c(Z_t)$. We postpone further discussion of $G^+_0(\vec r,\vec
r\,',\omega)$ to the subsection~\ref{subsec:bandaverage}. 

Finally, we define a dimensionless quantity as the normalized density
of the substrate states at a position $\vec R$ above the metal surface 
\begin{equation}
\label{eq:nudef}
\nu(\vec R,\omega)=e^{2Z/\lambda} \frac{\rho_s(\vec R,\omega)}{
\rho_s(\omega)}= - e^{2Z/\lambda} 
\frac{{\mathrm Im}\,G_0^+(\vec R,\vec R;\omega)}{\pi \rho_s(\omega)}.
\end{equation}
The tunneling current $I_0$ into a clean metal is
given by the first term in Eq.~(\ref{eq:Tt}). The current $I_0$ for
small bias $(V/\phi_s \ll 1)$ can be written with the above
definitions in a familiar form
\begin{eqnarray}
\label{eq:current0vsV}
&&I_0(\vec R_t,V)= \frac{2e}{h} \int_{-\infty}^\infty d\omega \times
\\ \nonumber 
&&\times 
\rho_t(\omega') \left [ f_t(\omega') - f_s(\omega)  \right ]
\rho_s(\omega) \pi t_c^2(\vec R_t) \nu(\vec R_t; \omega).
\end{eqnarray}
Here, $f_s(\omega)$ and $f_t(\omega)$ are the substrate and STM tip
Fermi functions, respectively, and $\omega'=\omega-eV$. The tip and
substrate are assumed to have common chemical potential
$\epsilon_{Fs}=\epsilon_{Ft}=0$ at zero bias $eV=0$ and we adopt the
convention of measuring the energies in the substrate-adsorbate
complex and in the tip from their respective Fermi levels at finite
bias. The bias $V$ is measured with respect to $\epsilon_{Fs}$ and is
defined as positive when the chemical potential of the tip
$\epsilon_{Ft}$ is raised. The functions $\rho_s$ and $\rho_t$ are the
substrate and tip densities of states, respectively. 

It follows from Eq.~(\ref{eq:current0vsV}) that the tunneling current
$I_0$ is independent of temperature if $\rho_t$, $\rho_s$, and
$\nu(\vec R_t,\omega)$ are independent of energy in the relevant energy
range. If $\rho_s$ shows structure on the scale of the temperature $T$
while $\rho_t$ is constant, the current will depend on the temperature
of the tip only, and vice versa. These statements are not limited to
the case of clean metal surfaces, but also hold for the substrate
with an impurity. The same is true for ${\mathcal G}$, the differential
conductance. 

The equilibrium current in the presence of the adsorbate is written by
expressing $\tilde \rho_{sat}$ with the notation and approximations
that lead to Eq.~(\ref{eq:current0vsV}). We define a modified matrix
element $\tilde t_a(\vec R_t, \vec R_0;\omega)$ for tunneling from tip
to the adsorbate state as  
\begin{equation}
\tilde t_a = t_a + \pi t_c \, \rho_s \, \Lambda\,V_a.
\end{equation}
The second term represents a coherent process of tip-to-surface
tunneling, through-surface-propagation, and surface-to-adsorbate
hopping. This is completely isomorphic with Fano's coupling of an
excited state (here the tip state) with the originally discrete state
``modified by admixture of states of the continuum''. The reader is
enthusiastically directed to the original Fano paper for further
enlightenment on this point. 

We introduce the Fano~\cite{Fano61PR} parameter $q(\vec  R_t, \vec
R_0;\omega)$ as 
\begin{equation}
q=\frac{\tilde t_a} {\pi t_c V_a \rho_s}.
\end{equation}
We will see later that this definition of $q$ makes the expression for
differential conductance formally equivalent with the Fano formula in
certain limits. It is rather straightforward now to evaluate 
$\tilde \rho_{sat}$ using~(\ref{eq:Tt}) and~(\ref{eq:tiplocaldensity})
in~(\ref{eq:currentfinalequilibrium}). After rearranging the terms, we
write the current $I_{eq}(\vec R_t, \vec R_0, \omega)$ in the presence
of the adsorbate resonance as
\begin{eqnarray}
\label{eq:currentvsV}
&&I_{eq}(\vec R_t, \vec R_0, V)=
\frac{2e}{h} \int_{-\infty}^\infty d\omega \; \rho_t(\omega')
\times \\ \nonumber
&&\times \left [ f_t(\omega') - f_s(\omega) \right ] 
\rho_s(\omega) \; \pi t_c^2(\vec R_t)\; Y(\vec R_t,\vec R_0,\omega),
\end{eqnarray}
with 
\begin{eqnarray}
\label{eq:Y}
Y=\nu +\sum_a\frac{\Gamma_{as}}{2}
\left \{ (\gamma^2-q^2)\; {\mathrm Im}\; G^R_a
+2q\gamma\; {\mathrm Re}\; G^R_a \right \}.
\end{eqnarray}
In our approximation, the localized nature of the tip and the
adsorbate enters through the position dependence of $t_a(\vec R_t, Z_0)$ 
and the substrate Green's function $G^+_0(\vec R_t,0;\omega)$.
The matrix element $t_a$ gives an exponentially decreasing amplitude
with increasing tip-adsorbate distance and the substrate Green's
function gives decreasing amplitude due to the phase difference
between electrons entering (or leaving) the surface at the adsorbate
site and leaving (or entering) at $(\vec R_\parallel,z=0)$ and also
due to the  exponential decay of the tip wavefunction with increasing
$k_\parallel$). We note that, in the wide band limit for the substrate
and with the tip near the surface above the adsorbate, $\tilde t_a
\approx t_a$, since in this limit ${\mathrm Re}\; G^+$ and thus
$\Lambda$ vanish. 
\begin{figure} 
\centerline{\epsfxsize=0.48\textwidth
\epsfbox{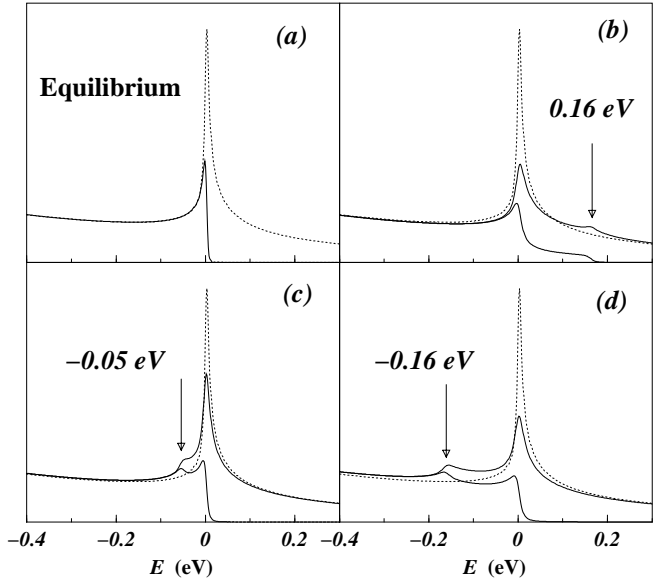}
}
\caption{The spectral function $\rho_a(\omega)$ and the occupied
density of states of a model Kondo system as a function of bias
voltage at small metal-tip separation given by
$\Gamma_{at}=0.1\Gamma_{as}$. (a) equilibrium, (b) - (d) finite
bias. Bold line = electron population, solid thin = spectral density,
dotted = equilibrium spectral density.
}
\label{fig:Kondospect}
\end{figure}
\subsection{Nonequilibrium effects at stronger tip-surface coupling} 
\label{subsec:nonequilibrium}

We now generalize Eq.~(\ref{eq:currentvsV}) for the equilibrium
tunneling current -- obtained in the lowest order in $t_{ap}$ and
$t_{kp}$ -- by including nonequilibrium effects. The general problem of
tunneling for arbitrary relative strength between the tunneling
amplitudes $t_{ap}$ and $t_{kp}$ and the hybridization matrix
$V_{ak}$ and for finite bias is formulated in Eq.~(\ref{eq:currentfinal}),
but the expression is quite complicated to evaluate in practice. In a 
typical STM experiment, the tunneling matrix elements $t_{ap}$ and $t_{kp}$
are much smaller than than $V_{ak}$. We can expect the nonequilibrium
effects to be important when, at small separations, the magnitude
of the two tunneling matrix elements is not a negligible fraction of
$|V_{ak}|$. However, we can always safely assume that $|t_{ak}|$,
$|t_{ap}|$ are smaller than $|V_{ak}|$ in the STM experiments under all
realistic conditions.

Therefore we make additional simplifications which are justified by these
relations. First of all, we replace $\tilde V_{ak}$
by $V_{ak}$ inside Eq.~(\ref{eq:Tpp'}) and (\ref{eq:selfenergy}). We 
neglect the modifications to the tip and substrate wavefunctions, i.e. 
replace $\tilde G_{kk'}$ by $\delta_{kk'}G_k^0$ and
$\tilde G_{pp'}$ by $\delta_{pp'}G_p^0$. We also neglect any deviations
from thermal electronic distribution in the substrate and tip, i.e. we
assume the validity of the fluctuation-dissipation theorem for the tip
and substrate Green's functions. On the other hand, when the tip-adsorbate
coupling is not negligible with respect to the adsorbate-metal hybridization,
the current into the resonance can be large enough to produce significant
nonequilibrium electronic population on the adsorbate since the time
scales for electron dissipation from the resonance into the metal and tip,
respectively, are comparable. In this case, the fluctuation-dissipation
theorem $G_a^< = f_s \rho_a$ is no longer valid for the adsorbate Green's
function and we must use the full nonequilibrium $G_a^<(\omega)$
instead of $f_s(\omega) \rho_a(\omega)$ in Eq.~\ref{eq:currentvsV}.

Under these assumptions, we find it convenient to write the total current
with the nonequilibrium effects as
$I_{tot}= I_{eq} + \delta I_{non}$, where $I_{eq}$ is formally given 
by~(\ref{eq:currentvsV}) and $\delta I_{non}$ is 
\begin{eqnarray}
\label{eq:currentnoneq}
\delta I_{non}&=&-\frac{2e}{h} \sum_a \int_{-\infty}^\infty d\omega \,
\pi^2 \rho_t \rho_s^2 t_c^2 V_0^2 \times \\ \nonumber 
&\times& ( f_s {\mathrm Im}G^R_a +\pi
G_a^< ) ( q^2 + \gamma^2 ),
\end{eqnarray}
where all adsorbate and substrate densities and Green's functions are
evaluated at energy $\omega$ and $\rho_t$ at
$\omega'=\omega-eV$. We omitted the spatial arguments for 
simplicity. The bias dependence enters through the self-consistent
solution of the adsorbate spectral density
$\rho_a(\omega)=-\frac{1}{\pi} {\mathrm Im} G^R_a(\omega)$ and the
``lesser'' Green's function $G_a^<(\omega)$. In the case of
noninteracting system, $(U=0)$, the spectral density does not depend
on the bias and the only nonequilibrium (finite bias) effect is given
by the difference between the equilibrium $G_{a,eq}^<(\omega)=f_s(\omega)
\rho_a(\omega)$ and the nonequilibrium density of occupied 
states $G_a^<(\omega)$, as featured in $\delta I_{non}$.

On the other hand, the spectral density $\rho_a(\omega)$ of Kondo
systems depends on the bias. This means that $I_{eq}$ also contains
nonequilibrium effects and is different from the equilibrium current
despite the subscript ``eq'' and its identical form. The effect of bias
on the spectral function depends on the tip hybridization with the
discrete impurity level and is similar to that of temperature for
$eV \leq T_K$ where it broadens the Kondo resonance. At larger biases
the broadening increases further and a second peak may develop at the
Fermi level of the tip, depending on the strength of the adsorbate-to-tip
hybridization $\Gamma_{at}=2 \pi \rho_t t_a^2$ compared to 
$\Gamma_{as}=2\pi \rho_s V_a^2$ for the relevant impurity
orbital.~\cite{WingreenMeir94PRB,SivanWingreen96PRB} 
In Fig.~\ref{fig:Kondospect}, we show for different bias voltages the
spectral function and electron occupation of the resonance for a model
Kondo system with $\Gamma_{at}$ equal to $\sim 10\%$ of
$\Gamma_{as}$ and under an additional assumption that $|t_{kp}| \ll |t_{ap}|$.
The model will be discussed in more detail in section~\ref{subsec:Kondo}.  

\subsection{Differential conductance in the limit of large tip-surface
separation}
\label{subsec:differential}

The differential conductance is obtained directly
from~(\ref{eq:currentvsV}) by differentiating it with respect to the
bias, i.e. ${\mathcal G}=dI/dV$. We do this here under the assumption
that the bias voltage is varied across a sufficiently narrow range so
that the density of tip states may be taken constant. Under these
assumptions the differential conductance ${\mathcal G}_{eq}$ is
\begin{eqnarray}
\label{eq:conductancevsV}
{\mathcal G}_{eq}(\vec R_t,\vec R_0,V)&=&\frac{2e^2}{h}
\int_{-\infty}^\infty d\omega \rho_t(\omega') \left
(-\frac{\partial f_t(\omega')}{\partial \omega} \right ) \\
\nonumber
&\times& \rho_s(\omega) \; \pi t_c^2(\vec R_t)\; Y(\vec R_t,\vec
R_0;\omega)  
\end{eqnarray}
and for the clean metal
\begin{eqnarray}
\label{eq:conductance0vsV}
{\mathcal G}_0(\vec R_t,V)&=& \frac{2e^2}{h} \int_{-\infty}^\infty d\omega 
\rho_t(\omega') \left (-\frac{\partial f_t(\omega')}{\partial \omega}
\right ) \times \\ \nonumber
&\times& \rho_s(\omega) \; \pi t_c^2(\vec R_t)\; \nu(\vec R_t,\omega).
\end{eqnarray}
These expressions neglect any changes to the tunneling barrier from
the finite bias voltage. When these approximation are not justified,
the conductance must be obtained by differentiating the expression for
current~(\ref{eq:currentvsV}) and~(\ref{eq:current0vsV}). This is
always the case for $I_{tot}$ of the previous section when
nonequilibrium effects are important.

It is known from the Anderson ``compensation
theorem''~\cite{Anderson61PR} that in the wide band limit
($\rho_s(\omega)$ is constant and unbounded), the presence of an
impurity does not affect the conduction electron density of states at
all. However, the density of conduction states is affected locally
even in this limit, as can be seen by setting $\psi_a=0$ in
Eq.~(\ref{eq:localdensity}). The perturbation of the conduction
electrons is probed by the STM directly if $t_a=0$, i.e. when
tunneling into the local state is absent due to either large
tip-adsorbate separation or because of symmetry. Under these
conditions if the STM conductance shows the signature of the local
resonance, it is a result of the perturbation of the LDOS of
conduction electrons. 

An important feature of the final result~(\ref{eq:currentvsV})
and~(\ref{eq:conductancevsV}) is that the role of the 
impurity resonance on the tunneling conductance is contained in the
Green's function $G_a$ of the local adsorbate state. It is then
possible to separate the problem into two steps. First, the solutions
for the adsorbate and substrate Green's functions are found for a
given system; then the tunneling conductance is calculated using the
solution in the expression~(\ref{eq:currentvsV})
or~(\ref{eq:conductancevsV}). Since the
equations~(\ref{eq:currentvsV}) and~(\ref{eq:conductancevsV}) have
been obtained under very general assumptions, they can be
used as a starting point in the study of a variety of tunneling problems
with appropriate approximations for the Green's functions and the
tunneling matrix elements. We demonstrate this in the next section,
where we first study noninteracting and then Kondo systems. If the
approximations to the tunneling matrix elements employed here are too
crude, the more general expression~(\ref{eq:currentfinal})
or~(\ref{eq:currentfinalequilibrium}) must be used.

We find that the lineshape $({\mathcal G}$ vs. $V)$ of the
adsorbate resonance depends sensitively on (1) the relative strength
of the tunneling matrix elements $t_{ap}$ and $t_{kp}$, (2) the
perturbation of the conduction electron states, (3) the lineshape of
the local resonance, i.e. the spectral function $-\frac{1}{\pi}\, {\mathrm
Im}\;G_a$ and ${\mathrm Re} \;G_a$, and on (4) temperature in the case
when $T \sim \Gamma$, the width of the resonance. Since the observed
perturbation of metal states and the tunneling matrix elements depend
on the tip position, so will the lineshape.

\subsection{The substrate Green's function $G^+_0$ and perturbation
of the conduction electrons: jellium surface} 
\label{subsec:bandaverage}

There are two ways in which the adsorbate state affects the tunneling
conductance: (A) direct tunneling into the discrete state; (B)
perturbation of the conduction electron states by the discrete state
which consequently contributes to the tip-to-continuum tunneling current. Both
contributions drop-off with increasing tip-adsorbate separation. The
direct tunneling into the resonance is controlled by $t_{ap}(\vec
R_t,Z_0)$ which is a function of the overlap between the tip and
adsorbate wavefunction and thus decays exponentially with the
distance. The perturbation of the continuum also vanishes at large
distances from the adsorbate. However, its spatial extent shows a more
complicated behavior and depends on the details of the electronic
structure of the substrate in resonance with the broadened discrete
state. It is anticipated that this contribution will show a
significantly longer range influence than the decaying exponential,
much in the spirit of Friedel oscillations. 

The position dependence of the perturbation enters through the Green's
function $G^+_0(\vec R_t,0;\omega)$. We note that the imaginary part
$\gamma$ appears explicitly in the expression for
conductance,~(\ref{eq:conductancevsV}), while the real part $\Lambda$
enters the definition of $\tilde t_a$. The STM-observable effects of
the spatially dependent perturbation of the conduction electrons
caused by the local impurity state are thus also controlled by
$G^+_0$~(\ref{eq:freemetalgreen}). In our formulation, a non-trivial 
$G^+_0$ is intimately related to the dependence of the tunneling
element $t_{kp}(\vec R_t)$ on the lateral tip position $\vec
R_\parallel$. With this in mind we focus on the specific problem 
of tunneling as a function of the tip-adsorbate separation.  

We consider a simple approximation for $G^+_0$ based on the
assumption that in the relevant surface region the surface
corrugations are smoothed out (jellium model) and both the Bloch
and/or surface state 
$\psi_k$~(\ref{eq:metalstate}) is given by
\begin{equation}
\label{eq:psik}
\psi_{\vec k}(\vec r)\propto e^{-\kappa_s z}
e^{i\vec k_\parallel\cdot\vec \rho}.
\end{equation}
The states with the smallest $\kappa_s$ have the longest tail into the  
vacuum region and thus will be the most important ones in the
tunneling process. These are the states with the smallest
$\epsilon_{k_\parallel}$. It is then reasonable to represent 
$\kappa_s$ in terms of the Taylor expansion around the minimum of
$\epsilon_{k_\parallel}$ with $\epsilon_k$ equal to the bias. In most
cases, it is reasonable to replace $\epsilon_k$ by its Fermi level
value. We expand $\epsilon_{k_\parallel}$ around its minimum as
$\epsilon_{k_\parallel}\approx k^2_\parallel/2m^*$, and write  
$\kappa_s =\lambda^{-1} + \lambda k^2_\parallel/2$ plus higher order
terms which we neglect. For states far from $\epsilon_{F_s}$ and at
small tip-surface separation, the expansion should be made around a
different value of $\lambda$. For the purpose of this paper, it is
sufficient to consider the Fermi level value $\lambda$. We then write
\begin{equation}
\label{eq:psifin}
\psi_{\vec k}(\vec r)\approx
e^{-z/\lambda}e^{-\lambda z k_\parallel^2 /2}
e^{i\vec k_\parallel\cdot\vec \rho}.
\end{equation}
As we will show later,
the second exponential $e^{-\lambda Z_t  k^2_\parallel/2}$ is a
measure of the tunneling current carrying $k_\parallel$, the property
that gives the STM tip its spatial resolution, and the third
exponential, $e^{i\vec k_\parallel\cdot\vec R_\parallel}$, controls
the dependence of the tunneling current on the lateral tip position.

With this approximation for Bloch states in the surface region
the substrate Green's function~(\ref{eq:freemetalgreen}) is
\begin{equation}
\label{eq:freemetalgreenapprox}
G^+_0(\vec R_t, 0;\omega) = e^{-Z_t/\lambda} \sum_k
\frac{e^{-\lambda Z_t k_\parallel^2 /2} e^{i\vec k_\parallel\cdot\vec
R_\parallel}|\psi_k(0)|^2}{\omega-\epsilon_k+i \eta }.
\end{equation}

For the bulk band state propagation, it is easy to show using
Eq.~(\ref{eq:gamma}) that
\begin{equation}
\label{eq:bessel}
\gamma(\vec R_t,\omega) = \int_0^1 dx J_0(k_\omega
R_\parallel\sqrt{1-x^2}) e^{-\lambda Z_t k_\omega^2(1-x^2)/2}
\end{equation}
and
\begin{equation}
\label{eq:bessel2}
\Lambda(\vec R_t,\omega) =\frac{1} {\pi \rho_s(\omega)}
{\mathcal P} \int_0^{2D} d\epsilon \rho_s(\epsilon)
\frac{\gamma(\vec R_t,\epsilon)}{\omega - \epsilon}, 
\end{equation}
where  $J_0$ is the zeroth order Bessel function and $k_\omega$ is the
wavevector of the substrate state of energy $\omega$. The normalized
density of (STM-accessible) conduction states $\nu(Z_t)$ a distance
$Z_t$ from the surface is 
\begin{equation}
\label{eq:bessel3}
\nu(Z_t,\omega) = \int_0^1 dx e^{-\lambda Z_t k_\omega^2(1-x^2)}.
\end{equation}
In calculating $\Lambda$, $\gamma$, and $\nu$ we assumed jellium-like
dispersion relation $\omega=k_\omega^2/2m^*$ and use parabolic density
of states $\rho_s(\omega)=1-\omega^2/D^2$. The incompatibility
of the density of states with the dispersion relation is not important
for the purpose of demonstrating the important band structure effects
at this level of simplification.  

Although the expressions~(\ref{eq:bessel})-(\ref{eq:bessel3}) are
valid for a very simple model of the surface, we believe they contain
the most important features of more realistic bulk electronic
structures. We now discuss these features beginning with $\gamma$,
Eq.~(\ref{eq:bessel}). At large $Z_t$, the dominant contribution to
the integral in $\gamma$ comes from small values of the argument $y$
in $J_0(y)$. In this case, we use the mean value theorem to
write~(\ref{eq:bessel}) as $\gamma(\vec R_t,\omega)=J_0(\bar k
R_\parallel) e^{-\lambda Z_t\bar k^2/2}$  where $\bar k=\alpha
k_\omega$ with $\alpha\in(0,1)$. Clearly, $\alpha \rightarrow 0$ as
$Z_t\rightarrow \infty$ and $\gamma(\vec R_t,\omega)$ is independent
of the lateral tip position. Since at the same time
$\Lambda\rightarrow 0$ and $t_a\rightarrow 0$, the STM has no spatial
resolution in this limit. As the tip moves closer to the surface the
spatial resolution increases. In the limit $Z_t=0$, the integral
in~(\ref{eq:bessel}) can be evaluated and  $\gamma(\vec
R_t,\omega)=j_0(k_\omega R_\parallel)$ where $j_0$ is the spherical
Bessel function of zeroth order.  

\begin{figure}
\centerline{\epsfxsize=0.48\textwidth
\epsfbox{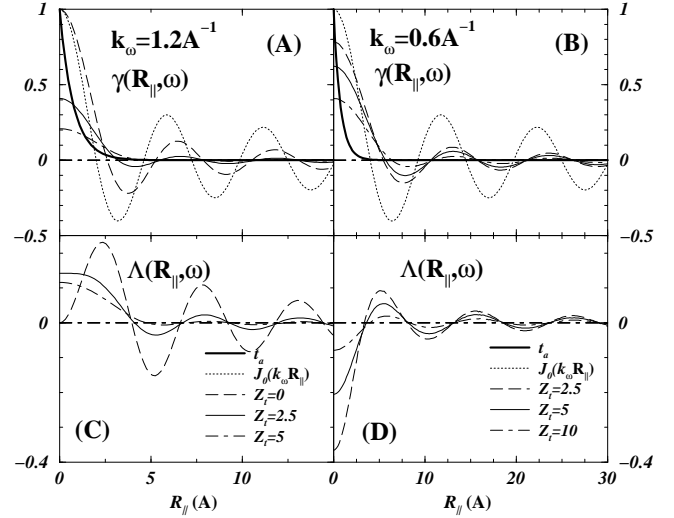}
}
\caption{
(A) and (B) show the typical spatial dependence of the tunneling
matrix element $t_a$ (bold) and of the substrate Green's function
$G^+$ for the jellium model evaluated at $k_\omega=1.2$~\AA$^{-1}$
(left panels) and $k_\omega=0.6$~\AA$^{-1}$ (right panels) for
different $Z_t$. We parameterize $t_a \propto e^{-R_\parallel/a}$ with 
$a=0.75$~\AA. Also shown is the Bessel function $J_0(k_\omega
R_\parallel)$ for the same $k_\omega$, i.e. $\gamma$ of a surface
state with $k_\omega$. The lower panels (C) and (D)
show $\Lambda$, the upper panels (A) and (B) show $\gamma$, $J_0$, and 
$t_a$. The comparison for the two wavevectors assumes identical
barrier (damping constant $\lambda$). 
}
\label{fig:2}
\end{figure}
The resonance lineshape depends on the ratio of the direct
tip-adsorbate tunneling amplitude ($t_a$) to the amplitude for
tunneling into the perturbed metal states ($t_c$) and on the
interference of conduction electrons scattering from the impurity.
The latter contribution is represented here by the substrate Green's
function $G^+$. Therefore, the presence of an impurity on the surface 
can be sensed spectroscopically even if the direct tunneling into the
resonance is negligible as is the case, for instance, of
$Ce/Ag(111)$~\cite{LietAl98PRL}. As our estimates for $t_a(R_\parallel)$
and $\gamma(R_t)$ indicate, the direct tunneling matrix element
$t_a(R_\parallel)$ falls off much more rapidly with $R_\parallel$ than
does $\gamma$. This is due to the limited spatial extent of the
tightly bound impurity orbital. Therefore, the relative importance of
tunneling into the perturbed continuum is likely to increase with the
lateral tip-adsorbate separation. 

Using the simple model for $G^+$, we show the typical length scales
in Fig.~\ref{fig:2}(A). We plot $t_a$ (bold solid line) parameterized
as $t_a \propto e^{-R_\parallel/\alpha}$ and normalized to one for
$R_\parallel=0$ together with $\gamma$ evaluated at three different
positions $Z_t$ above the surface and with
$k_\omega=1.2$~\AA$^{-1}$. The exponential fall-off for conduction
states at the Fermi level with work functions in the range $4-5$~eV
would be characterized by $\lambda=0.9$~\AA. The tightly bound
discrete state will have larger decay constant and we parameterize it
by $\alpha=0.75$~\AA. Clearly, $\gamma$ decays much slower than $t_a$ at
small tip-surface separations, but the difference in fall-off becomes
smaller with increasing $Z_t$. We also see by comparison with the
Bessel function $J_0(k_FR_\parallel)$ (light dotted) that the
spatial frequency decreases with increasing $Z_t$ and the oscillations
eventually disappear entirely. This is due to the fact that smaller
$k$-vectors have larger weight in the tunneling at greater $Z_t$ (see 
integral~(\ref{eq:bessel})). The real part $\Lambda$, shown in panel
(C) for the same k-vector, has a similar behavior. However, being a
Hilbert transform of the imaginary part, the nodes in $\Lambda$ appear
at the positions of local extrema of $\gamma$ and vice versa. As we
will see later, this property would lead to significant variations in
the lineshape with $R_\parallel$ if it survived in the real electronic
structure. Our results suggest that this is possible only at small
$Z_t$. We discuss the band structure effects in the following section.

The panels (B) and (D) show the same as (A) and (C) but for smaller
wavevector $k_\omega=0.6$~\AA$^{-1}$. Comparison between the right and
left thus demonstrates the strong dependence of the substrate Green's 
function on the wavevector itself, not just the product $k_\omega
R_\parallel$. The two most significant features are that with
decreasing $k_\omega$: (1) the frequency and damping of the
oscillations with $R_\parallel$ decrease and (2) the dependence on
$Z_t$ weakens. In comparing the two different energies, we assumed
that the damping constant $\lambda$ (i.e. the tunneling barrier) is
identical in the two cases. This would be the case in metals with
identical work functions for states at the Fermi energy, in one of 
which the bottom of the band were closer to the Fermi level (smaller
$k_\omega$). We note that $k_\omega=1.2$~\AA$^{-1}$ corresponds to
energy $\omega=5$~eV in the middle of the parabolic band with our
parameterization. Therefore, the value of $\Lambda(\vec R_t=0)=0$ at
this energy but is negative for smaller energies, e.g. for
$k_\omega=0.6$~\AA$^{-1}$, since in this case there are more high
energy continuum states repelling the discrete state downward than low
energy states pushing it up. We also see that the value of $\Lambda$
at $R_\parallel=0$ can change sign with $Z_t$ depending on the energy
$\omega$. We note that, since $\Lambda$ enters the expression for $q$,
the Fano parameter could also be negative and the asymmetry of the
resonance lineshape could be reversed.

\subsection{Electronic structure effects and the surface states on
(111) noble metals}  
\label{subsec:surfacebandaverage}

In the previous section we introduced a simple model of $G^+$ based on
the unperturbed jellium surface. In general, more realistic
behavior of $G^+$ can be obtained from electronic structure
calculations. Here we discuss qualitatively the electronic structure
effects with special attention to the (111) surfaces of noble metals
frequently used in STM studies.

It is well known that (111) surfaces of noble metals contain
Shockley surface states inside the projected two dimensional band gap
that forms on these
surfaces.~\cite{PlummerEberhadrdt82ACP,KevanGaylord87PRB}  
Both the surface state and bulk wavefunction are given by the same
general expression~(\ref{eq:metalstate}) outside of the metal
surface. However their overall degree of localization at the surface
is determined by the position of $\epsilon_{ss}$, the surface state 
eigenvalue, with respect to the band gap edges.  All other things
being equal, the most localized surface state occurs when $\epsilon_{ss}$
is at midgap. As $\epsilon_{ss}$ moves towards either band edge, the
extension of the evanescent oscillatory tail of the surface state wave
function into the bulk increases, ultimately becoming identical to a
periodic Bloch function when $\epsilon_{ss}$ hits the band edge. From
elementary normalization considerations, surface state extension into
the bulk and amplitude at the surface, as reflected in the scale
factor (or normalization constant) for the surface state tails (\ref{eq:psik})
extending into vacuum, are intimately related; greater population
within the bulk means lesser in the surface
region.~\cite{Gadzuk72JVST,KevanGaylord86PRL}
This surface state delocalization into the bulk allows
for the local density of bulk states at the surface to greatly exceed
that of the surface states, in which case the relative importance of
the surface state in the tunneling current will be small near the
surface.~\cite{Gadzuk72JVST} However, its importance increases with
increasing distance from the surface because the bulk states with
shorter wavefunction tails are eliminated from the tunneling. The
surface state accounts for about $50\%$ of the total signal in typical
STM tunnel junctions in
Au(111)~\cite{EversonJaklevic90JVacSciTech,Chenetal98PRL} and is known
to be responsible for the interference effects observed on these
surfaces near edges, impurities, and in quantum
corrals.~\cite{Crommieetal93SCI,Crommieetal95Phys,Burgi99PRL,HasegawaAvouris93PRL,AvourisLyo95CPL}  
It will also play a disproportionately important role in the resonance
tunneling at large lateral tip-adsorbate distance because its
contribution to $G_0^+$ does not decay as quickly as that for the bulk 
states. 

We see the different behavior of the bulk and the surface states in
the STM when we consider the propagator $G_0^+$ for the Shockley
state. This is again given by
Eq.~(\ref{eq:freemetalgreenapprox}). However, the k-sum now only
extends over the 2-D wavevector $k_\parallel$. Assuming parabolic
dispersion for the surface state, $\gamma$
is given by 
\begin{equation}
\label{eq:Shockleybessel}
\gamma(\vec R_t,\omega) = J_0(k_\omega R_\parallel) e^{-\lambda Z_t
k_\omega^2/2} 
\end{equation}
and
\begin{equation}
\label{eq:Shockleybessel2}
\Lambda(\vec R_t,\omega) = \frac{1}{\pi\rho_s(\omega)} \int_0^{2D}
d\epsilon \rho_s(\epsilon) \frac{\gamma(\vec R_t,\epsilon)}{\omega -
\epsilon}, 
\end{equation}
where, as before, $k_\omega$ is the 2-D wavevector of the substrate
state corresponding to energy $\omega$. The contribution of the
Shockley state to the normalized density of conduction states
$\nu(Z_t)$ is given by 
\begin{equation}
\label{eq:Shockleybessel3}
\nu(Z_t,\omega) = e^{-\lambda Z_t k_\omega^2}.
\end{equation}

The propagator $G_0^+$ for the surface state is essentially equal to
the Bessel function $J_0(k_\omega R_\parallel)$ weighted by the
exponential $e^{-\lambda Z_t k_\omega^2/2}$. Therefore the
oscillations are not damped with increasing $Z_t$ and only their
overall amplitude is diminished. Since the surface
state on the noble metal surfaces (111) have a short $k_F \sim
0.15-0.2$~\AA$^{-1}$, its propagator will have a much longer spatial
extent than that of the bulk states. The corresponding oscillations
thus have a spatial period of about 10 times that of the Bessel
function $J_0$ in Fig.~\ref{fig:2}(A) in agreement with the
experimental observation of Friedel oscillations. The contribution of
the surface state to the total current can carry information about an 
impurity on the surface over a long distance. The perturbation of the
surface state by the impurity should persist over several tens
of angstr\"oms. 

It is also known~\cite{EversonJaklevic90JVacSciTech} that the spectral
weight of the surface state decreases near surface imperfections. We
expect the same to be true near the adsorbate. While we have
explicitly taken into account the interaction of the conduction
states with the discrete state ``$a$'' through the adsorbate Green's
function $G_a$, all other adsorbate-metal interactions, such as
potential scattering of the conduction electrons from the adsorbate
and hybridization of the outer shell adsorbate electronic states
with the conduction electrons, are neglected in our model. In
principle, these ``residual'' adsorbate-metal interactions 
can be included by modifying $G^+$ and $t_{kp}$. Although a realistic
calculation of the system electronic structure is necessary to see the
effect of the adsorbate on the behavior of $G^+$ around the 
adsorbate, we believe that it will not produce oscillatory behavior in
$G^+$. In a typical metal,
several bands with anisotropic dispersion relations $\epsilon_{\vec
k}$ contribute to $G^+$ giving rise to more complicated behavior with
no single frequency. This will further reduce any oscillatory behavior 
seen in Fig.~\ref{fig:2}. At larger distance from the adsorbate, the
band structure of the clean surface will be reestablished and, as a
result, tunneling into the surface state.

Since the importance of the direct tunneling into the tightly bound
impurity orbital ``$a$'' relative to the tunneling into the metal
should be weak and decreases with increasing $R_\parallel$, it is
useful to study the asymptotic behavior of the conductance ${\mathcal
G}$ in the limit $t_a=0$. This is equivalent to replacing the Fano
parameter $q(\vec R_t, \vec R_0; \omega)$ by $\Lambda(\vec R_t;
\omega)$ inside $Y(\vec R_t, \vec R_0;  \omega)$,~(\ref{eq:Y}), in the
expression for conductance. It then follows that, if the oscillations
in $G^+$ persist, the lineshape should change with $R_\parallel$ and
antiresonances should form at positions where $\Lambda^2 >
\gamma^2$. Using $G^+=\pi\rho_s e^{-Z/\lambda}(\Lambda -i \gamma)$ and
\begin{eqnarray}
\label{eq:spatialinversion}
&&{\mathrm Im}\left \{ G^+(\vec R_t,\vec R_0)G^R_aG^+(\vec
R_0,\vec R_t) \right \}= \\ \nonumber
&&-\pi^2\rho_s^2 e^{-2Z_t/\lambda}\left \{ ( \gamma^2
- \Lambda^2 ) \, {\mathrm Im}\, G_a^R + 2 \Lambda \gamma
\, {\mathrm Re}\, G_a^R \right \}
\end{eqnarray}
we can write $\Delta {\mathcal G}_{eq}\equiv G_{eq}-G_0$ at zero temperature by
replacing (-$\partial f(\omega)/\partial \omega)$ by
$\delta(\omega-V)$ and using $t_c(Z)=t_0 e^{-Z/\lambda}$ as
\begin{eqnarray}
\label{eq:friedelosc2}
\Delta {\mathcal G}_{eq}(V) &=& - \frac{2e^2}{h} t_0^2 \rho_t(0)\, 
V_a^2 \times \\ \nonumber
&\times& {\mathrm Im}\left \{ G^+(\vec R_t,\vec R_0;V)\; G^R_a(V)
\; G^+(\vec R_0,\vec R_t;V) \right \}.
\end{eqnarray}
We see that the resonance in the conductance is a result of an
interference between different conduction states scattering resonantly
from the impurity. Its long range behavior on the (111) noble metal
surfaces is controlled by the surface states. Whether the resonance
can be observed at the large distances ($\ge 20$~\AA) depends on the
spectral weight of the surface state and on its hybridization $(\sim
V_a^2)$ with the impurity orbital ``$a$'' (usually $d$ or $f$).
Interesting spatial effects may be realized in system with suitable
boundary conditions. We believe that Eigler's quantum mirage of the
Kondo resonance inside the elliptical corral falls into this
category.~\cite{Manoharan00Nat} Based on the results of the previous
section, we do not expect ``Friedel'' oscillations at smaller
distances and with period of a few angstr\"oms characteristic of the
bulk k-vector, although we cannot completely rule these out for small
$Z_t$. 

\section{Discussion and examples}
\setcounter{equation}{0}
\label{sec:discussion}

The equations~(\ref{eq:currentvsV}) and~(\ref{eq:conductancevsV}) were
derived under rather general assumptions. They are suitable as
a starting point for numerical investigations given the necessary
input from electronic structure calculations. In the rest of the
paper, we discuss the implications of our theory for several specific
cases of interest. In all of these cases we use our simple model for
$G^+$ based on the jellium surface and the DOS given by
$\rho_s(\omega)=\rho_t(\omega)=1-\omega^2/D^2$ with $D=5$~eV the band
half width. In order to eliminate the exponential fall-off in the
tunneling conductance with the tip-surface separation and the
background distortions, we plot the normalized change in conductance
due to the additional impurity defined as 
\begin{equation}
\Delta {\mathcal G}_{eq}(V)
\equiv h ({\mathcal G}_{eq}(V)-{\mathcal G}_0(V))/(2\pi e^2\rho_t(0)
t_c^2),
\end{equation}
where ${\mathcal G}_{eq}$ is given by~(\ref{eq:conductancevsV}) and
${\mathcal G}_0$ by~(\ref{eq:conductance0vsV}). This is equivalent to
replacing $Y$ by $\Delta Y = Y -\nu$ in the expression for ${\mathcal
G}$. 

Although we were motivated by the experimental observation of the
Kondo resonance~\cite{MadhavanetAl98S,LietAl98PRL}
and this work is mostly applied to the tunneling through the Kondo
resonance, we discuss many of the tunneling properties on the simple
noninteracting model.  We do this primarily because most of the STM
observable characteristics of the Fano resonance are common to the
single particle and Kondo resonances, despite the difference in
processes that give rise to the two resonances. We wish to point
out these general features on a model that is conceptually far simpler
and more familiar to the surface science community than the Kondo 
model, and emphasize that the resonances can also be observed
in systems with nonmagnetic impurities with a tightly bound orbital
near the Fermi level. Finally, the connection with Fano result and the 
consequences of the spatial resolution of the STM become more
transparent when the same noninteracting Anderson Hamiltonian is
used. 

\subsection{Noninteracting adsorbate}
\label{subsec:nonint}

We begin our discussion with an adsorbate-metal system
described by the non-interacting Anderson model ($U=0$). The impurity
resonance is characterized by its energy $\epsilon_0$ and the width
$\Gamma_{as}$. The retarded Green's function $G^R_a(\omega)$ for the
adsorbate state is 
\begin{equation}
\label{eq:U0resonancedensity}
G^R_a = (\omega-\epsilon_0 - {\mathrm Re}\, \Sigma_a + i
\Gamma_{as}/2)^{-1} 
\end{equation}
where $\Gamma_{as}(Z_0,\omega)$ is defined in Eq.~(\ref{eq:gamma1})
and ${\mathrm Re}\;\Sigma_a(Z_0,\omega) ={\mathcal P}\sum_k
|V_{ak}(Z_0)|^2(\omega-\epsilon_k)^{-1}$ is the real part of the  
self-energy for the noninteracting Anderson model (not to be confused
with the real part of the substrate Green's function $\Lambda$ given
in Eq.~(\ref{eq:Lambda})). Following Fano we now define the
dimensionless energy parameter $\epsilon(\vec R_0,\omega)$ by
\begin{equation}
\label{eq:renormalizedlevel}
\epsilon=\frac{2 (\omega-\epsilon_0-{\mathrm Re}\;\Sigma_a
)} {\Gamma_{as}}.
\end{equation}

We neglect all nonequilibrium effects since they are likely to be
insignificant for the noninteracting system under most experimentally
realizable conditions. The differential conductance, in
lowest order in $t_c$ and $t_a$, is given by
Eq.~(\ref{eq:conductancevsV}) where $Y$ for the noninteracting
system takes the form
\begin{equation}
\label{eq:conductancevsVU0}
Y \equiv Y_0 =
\nu + \frac{q^2-\gamma^2 +
2\epsilon\gamma q} {1+\epsilon^2}.
\end{equation}
All terms are evaluated at energy $\omega$ and at the 
appropriate  tip position. We note that $Y_0(0,0,\omega)\equiv Y_{00}$
characterizing the unphysical case of the STM tip in contact with
surface at the position of the adsorbate (embedded in the surface)
has the analytic form obtained by Fano 
\begin{equation}
\label{eq:Yfano}
Y_{00} =\frac{(q+\epsilon)^2} {1+\epsilon^2},
\end{equation}
although the inherent energy dependence of $q$ (through
$\Lambda(\omega)$) could distort the pure Fano character of the
lineshape, even for this ``almost atomic physics'' STM example.
At all other tip positions, the shape of the resonance will be
described by the more general expression~(\ref{eq:conductancevsVU0}),
which is equivalent to a sum of a ``Lorentzian'' plus a ``Fano''
profile,
\begin{equation}
Y_0=\frac{\nu'} {1+\epsilon^2} + 
\nu \frac{(q'+\epsilon)^2} {1+\epsilon^2}
\end{equation}
with $\nu'\equiv \nu + q^2 - \gamma^2 -q^2\gamma^2/\nu$ and $q'\equiv
q\gamma/\nu$. Decomposition of $Y_0$ in this manner may be useful when
analyzing experimental lineshapes. Note that the additional intrinsic
energy dependences of $\nu$ and $\gamma$ could further distort the
standard lineshape.
\begin{figure}
\centerline{\epsfxsize=0.48\textwidth
\epsfbox{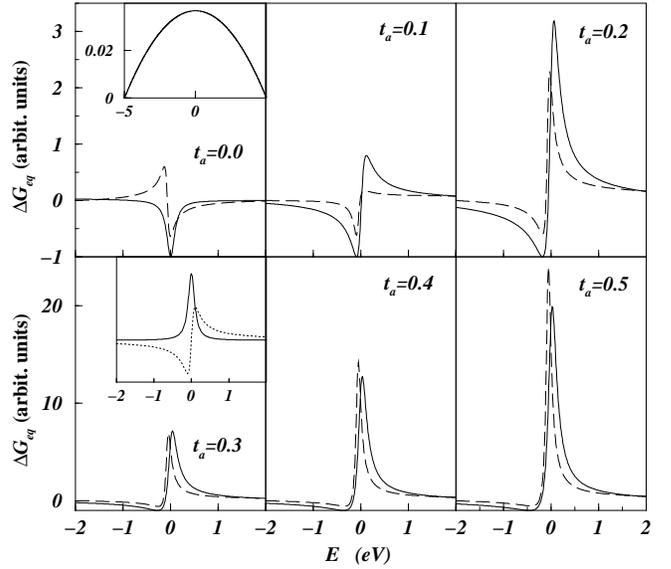}
}
\caption{Differential conductance $\Delta {\mathcal G}_{eq}$ as a
function of the strength of the direct tunneling matrix $t_a$ (in
units of $t_c$). The
model parameterization is given in text. The two curves correspond to
two different impurity level energies $\epsilon_0=5$~eV (solid) 
and $2$~eV (dashed) from the bottom of a symmetric band $10$~eV wide.
The two insets show the model density of states $\rho_s$ for the
conduction electrons (upper panel) and the imaginary (solid) and real
(dashed) part of $G_a(\omega)$ (lower panel).
}
\label{fig:3}
\end{figure}

\subsubsection{Lineshape dependence on electronic structure and on the 
relative strength of $t_a$ and $t_c$}

We first show (Fig.~\ref{fig:3}) the dependence of $\Delta {\mathcal
G}_{eq}$ on the ratio $(t_a/t_c)$ for $\vec R_t=0$ and a resonance at the
Fermi level. The solid line corresponds to a resonance at the
center of a parabolic band (symmetric around its center),
i.e. $\epsilon_0=5$~eV from the bottom of the band and the dashed line
corresponds to a resonance at $\epsilon_0=2$~eV from the bottom of the
band. The two energies correspond to the band energy $\epsilon_k$ with
wavevector $k=1.2$~\AA$^{-1}$ and $k=0.6$~\AA$^{-1}$ in our jellium
model (Fig.~\ref{fig:2}). The resonance width is $\Gamma_{as}=0.2$~eV
in both cases. At zero temperature, from Eq.~(\ref{eq:conductancevsV})
and~(\ref{eq:conductancevsVU0}) we write
\begin{equation}
{\mathcal G}_{eq}(V)=\frac{2e^2}{h} \rho_s(V) \pi \rho_t(0) t_c^2
Y_0(V)
\end{equation}
In order to make connection with the Fano result, we plot the
conductance for small tip-metal separation with the tip above the
adsorbate ($\vec R_t=0$). The lineshape $Y_0$ is then given by the
Fano formula~(\ref{eq:Yfano}). The plotted quantity $\Delta {\mathcal
G}_{eq}(V)$ in Fig.~\ref{fig:3} is then given by
\begin{equation}
\Delta {\mathcal G}_{eq}(V)=\rho_s(V) 
\left ( \frac{(q+\epsilon)^2} {1+\epsilon^2} - 1 \right ).
\end{equation}

The Fano parameter $q$ depends not only on the ratio $(t_a/t_c)$ but
also on energy and electronic structure. We see this most clearly in
the first panel where 
$t_a=0$. The resonance placed at the center of the band
produces a symmetric dip in $\Delta {\mathcal G}$ characteristic of
$q=0$, whereas the resonance at $\epsilon_0=2$~eV has an asymmetric
lineshape due to the negative contribution from $\Lambda$ to $q$ (see
Fig.~\ref{fig:2}). Its lineshape actually becomes symmetric at finite
value of $t_a$. The value of $t_a$ inside each panel is given in units
of $t_c$. The inset in the upper panel shows the model density of 
conduction states $\rho_s$ and the lower panel inset shows the
spectral function $\rho_a=-\frac{1}{\pi} {\mathrm Im} \, G_a^R$
(solid) and ${\mathrm Re}\, G_a^R/\pi$ (dotted) for the level at the
center of the band. 

As the strength of the direct tunneling $t_a$ increases with respect 
to $t_c$, the resonance develops its characteristic asymmetric shape
and, eventually, at large $t_a/t_c \gg 1$ it acquires the shape
nearly indistinguishable from that of the impurity spectral function
$\rho_a(\omega)$. With increasing tip-adsorbate separation, the signal
from the resonance must disappear as both the tunneling element $t_a$
and $G^+(\vec R_t,\omega)$ tend to zero. The differential conductance
is then determined by the density of states of the clean surface. This
property is not present in the Fano expression. We now discuss this
behavior.

\subsubsection{Lineshape dependence on the tip-surface separation}

\begin{figure}
\centerline{\epsfxsize=0.48\textwidth
\epsfbox{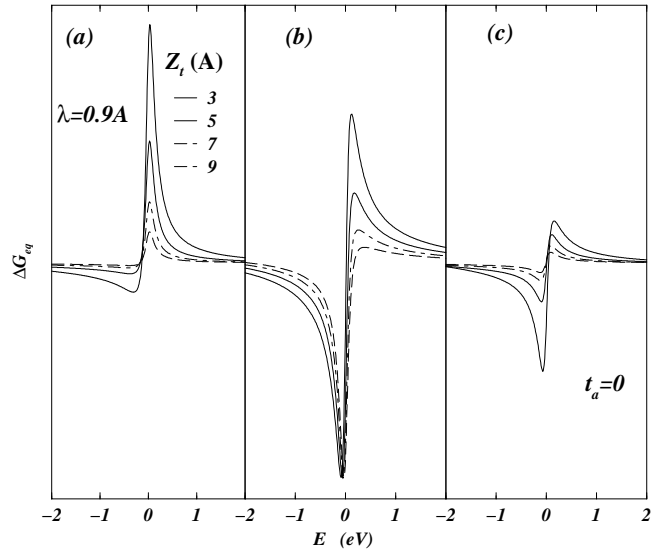}
}
\caption{ Differential conductance $\Delta {\mathcal
G}_{eq}$ vs. $Z_t$ for a model described in the text. (a) the complete
dependence on the tip distance $Z_t$; (b) assumes $Z_t=0$ inside $G^+$
and thus neglects the wavevector dependent effects; (c) is same as (a)
but with $t_a=0$, i.e. it only includes the effect of decreasing
spatial sensitivity incorporated through the substrate Green's
function $G^+$. The vertical scale is arbitrary but identical in all
panels. 
}
\label{fig:4}
\end{figure}

Using the same model system as in the previous section with the 
resonance at the center of the band ($\epsilon_0=5$~eV from the bottom
of the band), we demonstrate the dependence on $Z_t$ (with $\vec
R_\parallel = 0$) in Fig.~\ref{fig:4}. We make the following model for
the tunneling matrix element $t_a(\vec R_t, \vec R_0)$ and $t_c(\vec
R_t)$. The exponential fall-off of the metal and adsorbate
wavefunctions is controlled by different decay constants. The
adsorbate state $\psi_a$ is tightly bound especially for narrow
resonances of interest here. The conduction electron wavefunctions, on
the other hand, typically belong to the outer $s$ or $p$ orbitals and
have longer tails into the vacuum. As a consequence, the ratio
$t_a/t_c$, and thus also the Fano parameter $q$, changes with
$Z_t$. In order to incorporate this property, we use the matrix
elements~(\ref{eq:tdp2}) and $t_c = t_0 e^{- Z_t/\lambda}$, where
$\alpha=0.75$~{\AA}, $\lambda\approx 0.9$~{\AA}, $t_0=25$~meV, and
$t_a=0.1t_c$ at $Z_t=2$~\AA. Under these conditions, the $q$ parameter
tends to zero with increasing $Z_t$. 

The panel (a) shows the normalized $\Delta{\mathcal G}_{eq}$ for this 
model. The lineshape undergoes only moderate changes with $Z_t$ within
the experimentally relevant range. We expect this to be a general
property. In order to understand the behavior, we discuss the
lineshape dependence on tip-surface separation conceptually in terms
of two contributions: (1) different decay constants for the discrete
$\psi_a$ and metal $\psi_k$ states at the Fermi level, and (2)
different decay constant for metal states at $\epsilon_{Fs}$ with 
different $k_\parallel$. We separate the observable consequences of
these two effects in panels (b) and (c). The first contribution
produces changes in $q$ due to the changing relative strength between
$t_a$ and $t_c$. We demonstrate this in Fig.~\ref{fig:4}(b) where only
this contribution is taken into account by setting $Z_t=0$ inside the
substrate Green's function $G^+$, or equivalently by setting
$\gamma=\nu=1$ and $\Lambda=0$ in
Eq.~(\ref{eq:conductancevsVU0}). This limit does not correspond to a
real situation and does not lead to the correct $Z_t \rightarrow
\infty$ limit. It is shown here merely as an example of the
contribution (1) to the $Z_t$ dependence of the tunneling
conductance. With our parameterization, this case is identical 
with $q$ changing from $q\approx 0.8$ at $Z_t=3$~\AA \, to $q\approx
0.2$ at $Z_t=9$~\AA. As $Z_t$ increases further, $q\rightarrow 0$ and the
resonance becomes symmetric. However, we see that the normalized
conductance $\Delta {\mathcal G}_{eq}$ does not vanish in the limit
$Z_t\rightarrow \infty$. 

Fig~\ref{fig:4}(c) takes the second contribution (2) into account
while leaving out the first one. We chose $t_a/t_c=0=$ constant, which 
would be the case if there were no direct tunneling into the discrete 
state. In this extreme limit, any changes in lineshape are a
consequence of the varying weight that different $k_\parallel$ metal
states play in the tunneling at different $Z_t$. This occurs because
the $k_\parallel=0$ metal wavefunctions given by Eq.~(\ref{eq:psifin})
have the greatest extension into the vacuum
and as a result the spatial resolution of the tip decreases. Therefore
the signature of the resonance in ${\mathcal G}$ decreases even after
normalization of the current for different $Z_t$ as the ratio $\Delta
Y/ \nu \rightarrow 0$ with $Z_t\rightarrow \infty$. Fig~\ref{fig:4}(a)
shows the combined effect of the contribution and represents realistic
conditions. It accounts correctly for the  changing lineshape, as well
as its disappearance.  We again emphasize that realistic band
structure is desirable for making quantitative statements.

Finally, we comment on the experimental issues. It is clear that
the lineshape dependence on $Z_t$ will be observable only if it can be
studied over a reasonably large range of $Z_t$, this being limited by
the experimental resolution and detection capabilities. The most
favorable case is one in which the direct tunneling $t_a$ into the
resonance is strong at small $Z_t$, i.e. $q$ is large, and the
difference in relevant decay constants for the adsorbate and metal
electrons, $a$ and $\lambda$ respectively, is large. This is not the
case in the experiments~\cite{MadhavanetAl98S,LietAl98PRL} where $q$
is small. Therefore, we do not expect significant changes in the 
lineshape with $Z_t$ in these systems. Since our model is based on
realistic parameterization, we expect the behavior shown in 
Fig.~\ref{fig:4} to serve as a guide for order of magnitude
estimates for the spectral dependence on $Z_t$.

The direct effect of the STM tip on the system and thus also on the
lineshapes is not taken into account here. This issue is discussed in
section~\ref{subsubsec:Kondononeqcurrent}.

\subsubsection{Lineshape dependence on the lateral tip position}
\label{subsubsec:lateral}

\begin{figure}
\centerline{\epsfxsize=0.48\textwidth
\epsfbox{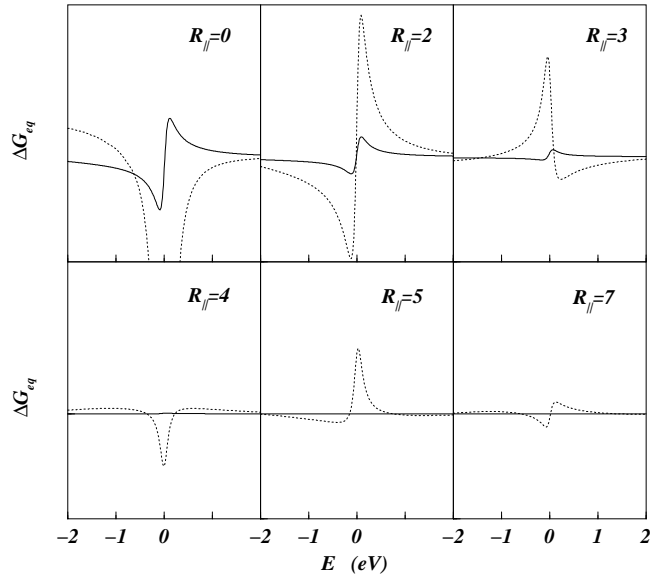}
}
\caption{ The normalized differential conductance $\Delta {\mathcal
G}_{eq}$ as a function of the lateral tip position  $R_\parallel$ for
the same model as in Fig.~\ref{fig:4}: $Z_t=5$~\AA (solid) and
$Z_t=0$~\AA (dotted). The contribution of the surface states to
$G_0^+$ is not included here.
}
\label{fig:5}
\end{figure}

As we already discussed in section~\ref{subsec:bandaverage},
the resonance lineshape depends on the relation between the spatial
dependence of the direct tunneling and the propagation of the
adsorbate-induced perturbation through the metal. Fig.~\ref{fig:2}
shows that the direct tunneling into the resonance $(t_a)$ is expected
to fall off faster than the perturbation. Therefore, at large
$R_\parallel$, the lineshape will be given by the $t_a=0$ limit
conductance. We show the dependence of
$\Delta {\mathcal G}_{eq}$ on $R_\parallel$ in Fig.~\ref{fig:5}. We do
this again for the model described in the previous section
(Fig.~\ref{fig:4}) with $t_a=0$ and a resonance at the center of the
band, i.e. $\epsilon_0=5$~eV from the bottom of the band which
corresponds to $k_\omega=1.2$~{\AA}$^{-1}$ in Fig.~\ref{fig:2}. The
solid line corresponds to $Z_t=5$~{\AA} and the dashed line to
$Z_t=0$~\AA. 

The unphysical case of $Z_t=0$ (dashed) is shown to emphasize the
possible consequences of the oscillations in $G_0^+$ displayed in
Fig.~\ref{fig:2}. We chose the lateral tip positions in the figure to
coincide with the nodes and zeros of $\Lambda$ and $\gamma$ to show
the dramatic changes in the lineshape with $R_\parallel$ due to the
oscillations in $\Lambda$ and $\gamma$. Since the spatial decay of the
oscillations is small at $Z_t=0$, the sequence of resonances and
antiresonances appear in the range $R_\parallel\in(0,10)$~{\AA}. The
possibility for such antiresonances is discussed implicitly in the
work of Kawasaka~\cite{Kawasakaetal99JAP} and explicitly by Schiller
and Hershfield~\cite{SchillerHershfield00PRB}. However, this behavior
is not observed in the experiments by Madhavan {\em et 
al.}~\cite{MadhavanetAl98S} and Li {\em et al.}~\cite{LietAl98PRL} due
to the smoothing of the electronic structure with increasing distance
from the surface that we discussed in~\ref{subsec:bandaverage}. 

In fact we would not expect the dramatic variations in line shape with
$R_\parallel$ reported by Schiller and
Hershfield~\cite{SchillerHershfield00PRB} to be observed. The reason is
apparent from the behavior of $G^+$ as a function of $Z_t$
(Fig.~\ref{fig:2}). As the tip distance from the surface increases the
oscillations are destroyed by the increasing weight of the lower
frequency (small $k_\parallel$) components at larger $Z_t$ interfering
destructively with those given by $k_\omega$. For this value of
$k_\omega$, the oscillations are effectively damped when $Z_t\ge
5$~{\AA} and the shape of the resonance does not change significantly
as shown by the bold line in Fig.~\ref{fig:5}. We expect that band
structure effects will suppress the oscillations even further.

We also find that the spatial extent of the resonance in the spectrum
should decrease as the STM is retracted, as long as the signal is due
to the bulk states. At $Z_t=0$~\AA, the resonance
is still visible at $R_\parallel \sim 10$~{\AA} but only to about
$R_\parallel \sim 4$~{\AA} at $Z_t=5$~\AA. This is a somewhat shorter
distance than that found experimentally for $Co/Au(111)$ and
$Ce/Ag(111)$~\cite{MadhavanetAl98S,LietAl98PRL}. Although the Fermi
wavevector $k_F\approx 1.2$~\AA$^{-1}$ used in Fig.~\ref{fig:5} is
close to the free electron value of $k_F$ for the noble metals, the
disagreement is not surprising since we made no real attempt at
realistic electronic structure description. Smaller values of $k_F$
would increase the spatial extent as would smaller values of $Z_t$ and
$\lambda$.

Interestingly, the $Z_t=0$ [dotted] lineshape progression
shown in Fig.~\ref{fig:5} is qualitatively similar to the family of
lineshapes that would be expected from surface state propagation, but
with $R_\parallel$, the lateral tip-adatom separation rescaled upward
by nearly an order of magnitude. This claim is based on the qualitative
similarity between the bulk $G^+_0$ at $Z_t=0$ and the surface
$G^+_0$. The bulk $\gamma(R_\parallel, \epsilon_F) = j_0(k_F R_\parallel)$
[the dashed curve in Fig.~\ref{fig:2}] at $Z_t=0$ the analogous
surface state $\gamma(R_\parallel, \epsilon_F)= J_0(k_F R_\parallel)$ [dotted
curve in Fig.~\ref{fig:2}] both exhibit long range oscillations
unlike the bulk state at $Z_t \ge 5$\AA.  However since $k_F \sim
(0.1-0.2)$\AA$^{-1}$ for the surface state band, $J_0(k_F
R_\parallel)$ shown in Fig.~\ref{fig:2}(A) for $k_F=1.2$\AA$^{-1}$
should be plotted with this smaller $k_F$ when referring to actual
noble metal surface state bands, in which case the observable
$R_\parallel$-dependent lineshape evolution in Fig.~\ref{fig:5}  
would still be representative, but with $R_\parallel$ rescaled by the
factor $1.2/0.15 =8$. From this it is easy to appreciate that the
dramatic lineshape variations will occur mainly at very large lateral
separations. Clearly, realistic electronic structure calculations are
necessary to answer the more quantitative questions. 

At large values of $R_\parallel$ and for broader resonances, 
an additional mechanism for distortion of the lineshape is possible
if the relative change in the length of $k_\omega$ in the energy
range given by the resonance width $(\Gamma)$ near $\epsilon_0$ is
large. In principle, this gives rise to the possibility of the
oscillatory behavior of $J_0$ with respect to energy exhibiting itself 
in the spectrum. However, the conditions for this effect in ${\mathcal
G}(\omega)$ would require the relative change $\Delta
k_\omega/k_\epsilon \approx \sqrt{1+\Gamma/2\epsilon_0} -
\sqrt{1-\Gamma/2\epsilon_0}$ to be $> 2\pi/k_\epsilon
R_\parallel$. In this case, the argument of the Bessel function
in~(\ref{eq:bessel}) will vary over several periods starting at a
small value at the bottom of the resonance. It is however clear that
the width of the resonance and the distance from the impurity would
have to be much larger than in the recent experiments with Kondo
impurities. Moreover, these oscillations could only be observed
through the surface state due to the damping of oscillations in the
bulk. 
\begin{figure} 
\centerline{\epsfxsize=0.48\textwidth
\epsfbox{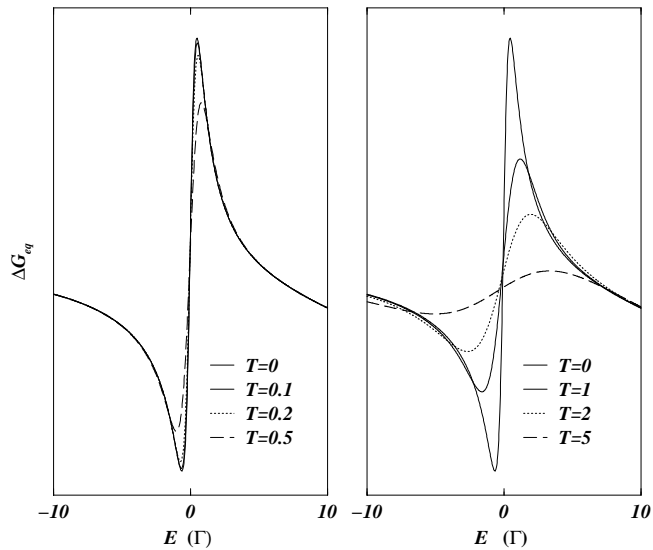}
}
\caption{  $\Delta {\mathcal G}_{eq}$ as a function of temperature for 
$Z_t=5$~\AA\, and the same system as in Fig~\ref{fig:4}. Temperature
and energy are given in units of the resonance width, $\Gamma$.
}
\label{fig:6}
\end{figure}

\subsubsection{Temperature dependence of differential conductance}

The temperature dependence of the tunneling conductance through a
narrow resonance is an important issue especially in considerations of
tunneling through the Kondo resonance which itself is temperature
dependent. We revisit this issue in the
section~\ref{subsubsec:Kondoequilibrium} on Kondo effect. Here, we
demonstrate the effect of the broadening in the Fermi function on the
spectrum when the impurity spectral function is independent of
temperature.

We assume the density of tip states and the substrate conduction
electrons to be constant on the scale of the width $\Gamma$
near the tip Fermi level and around the resonance. The
temperature dependence of the conductance then follows from
equation~(\ref{eq:conductancevsV})
with~(\ref{eq:conductancevsVU0}). We consider only the case of the tip
directly above the adsorbate for simplicity and write
\begin{equation}
\label{eq:conductancevsVU0temp}
{\mathcal G}(V)=\frac{2e^2}{h} \pi \rho_t \rho_s \int_{-\infty}^\infty
d\omega \left ( - \frac{\partial f_t(\omega')}{\partial \omega} 
\right ) t_c^2 Y_0(\vec R_t,\omega). 
\end{equation}

It follows from this expression that the differential conductance only 
depends on the temperature of the STM tip. This is a consequence of
the assumption that the density of tip states is constant in the
relevant energy range and the tunneling barrier is not modified by the
bias. The differential conductance is then a
function of the spectral density of the substrate states and is
independent of their occupation. Generally, if the DOS in the STM tip
varies significantly on energies $\sim \Gamma_{as}$, the substrate
temperature would also enter. If the density of states in both
the substrate and the tip were constant, no temperature dependence
would be observed. It follows from Eq.~(\ref{eq:conductancevsVU0temp})
that the temperature $T$ must be $\sim \Gamma$, in order to 
have a significant effect on the conductance. 

We show the temperature dependent ${\mathcal G}_{eq}$ in
Fig.~\ref{fig:6}. The spectral density $\rho_a(\omega)$ with the
narrow resonance at the Fermi level (the same system parameters as in
Fig.~\ref{fig:3}) is independent of $T$. The STM tip is directly above
the resonance with $Z_t=5$~{\AA}. In the range $T \leq 0.1 \; \Gamma$,
no temperature dependence is noticeable. However, when $T \sim
(0.2-0.5) \; \Gamma_{as}$ the differential conductance shows a rather
strong dependence on temperature, and at $T \gg \Gamma$ the
sensitivity of the STM to the resonance disappears. The temperature
dependence in Fig.~\ref{fig:6} comes entirely from the broadening of
the Fermi function of the tip. 

\subsection{Tunneling into Kondo resonance}
\label{subsec:Kondo}

In the previous section~\ref{subsec:nonint}, we discussed the STM
conductance in tunneling through a noninteracting impurity [$U=0$
in Hamiltonian~(\ref{eq:SystemHamiltonian})], frequently referred to
as the resonant level model (RLM). We now turn to the case of magnetic
impurities and tunneling through a Kondo resonance. We begin with the
case of a weak tip-metal coupling. However, for the Kondo systems this
assumption is more restrictive than for the RLM model, and for this 
reason, we later take advantage of our nonequilibrium approach to
account for the direct effect of the tip on the impurity spectral
density, while still neglecting the tip's effect on the metal states. 

\subsubsection{Conceptual and theoretical approach}
\label{subsubsec:conceptual}

Since our earlier derivation of the current and
conductance is valid for arbitrary interaction [$U \neq 0$ in 
Eq.~(\ref{eq:SystemHamiltonian})], the final
results~(\ref{eq:currentvsV}) and~(\ref{eq:conductancevsV}) also hold
in the Kondo and mixed-valent regimes of the Anderson model (i.e. $U
\gg \Delta$). The properties of the adsorbate enter through the
Green's function $G_a$. The problem is thus reduced to finding the one
electron Green's function $G_a$. 

\begin{figure} 
\centerline{\epsfxsize=0.48\textwidth
\epsfbox{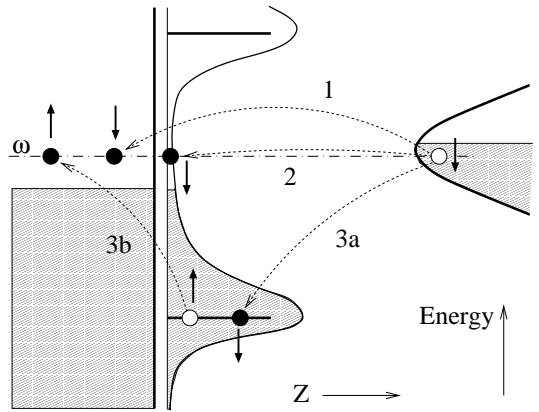}
}
\caption{ Possible scattering channels for an electron tunneling from
tip to metal through a magnetic impurity adsorbed on the surface.
}
\label{fig:7}
\end{figure}

However, we first consider the tunneling for a spin $1/2$
($a\equiv\sigma$) impurity in the Kondo limit,
$(\epsilon_{Fs}-\epsilon_0) \gg \Gamma$ and
$(\epsilon_0-\epsilon_{Fs}+U)\gg \Gamma$. The Kondo resonance has a
very small weight and is due to spin fluctuations. The possible
tunneling channels in this case are shown in Fig.~\ref{fig:7} as
processes (1) and (3). The system can be described by the Kondo
Hamiltonian in this limit 
\begin{eqnarray}
\label{eq:KondoHamiltonian}
H_s(Z_0)&=&\sum_{k\sigma}\epsilon_k c^\dagger_{k\sigma}c_{k\sigma}+
\sum_{p\sigma} \epsilon_p c^\dagger_{p\sigma} c_{p\sigma}+\\ \nonumber 
&+& \sum_{kp\sigma} \{ t_{kp}(\vec
R_t)c^\dagger_{k\sigma}c_{p\sigma}+{\mathrm H.c.} \}+\\ \nonumber
&+&J_s\sum_{kk'\sigma\sigma'} ( c^\dagger_{k\sigma}\vec s_{\sigma\sigma'}
c_{k'\sigma'} )\cdot \vec S + \\ \nonumber &+& 
J_t\sum_{pp'\sigma\sigma'} ( c^\dagger_{p\sigma}\vec s_{\sigma\sigma'}
c_{p'\sigma'} )\cdot \vec S + 
\\ \nonumber &+& J_{st} 
\sum_{kp\sigma\sigma'} \{ (c^\dagger_{k\sigma}\vec s_{\sigma\sigma'}
c_{p\sigma'} )\cdot \vec S + H.c. \}
\end{eqnarray}
where the first three terms were also present in the total Hamiltonian 
introduced in section~\ref{sec:modelandapprox} and describe the
unperturbed metal and tip states and the coupling between the two. The
remaining terms give rise to spin fluctuations in the presence of the
magnetic impurity. The terms with couplings $J_s$ and $J_t$ correspond
to the exchange interaction of the local spin with the substrate and
tip electrons, respectively. The last term ($J_{st}$) corresponds to
the effective tip-substrate exchange interaction in which charge is
transported between the tip and the surface. This Hamiltonian can be
obtained from $H_{tot}$ of section~\ref{sec:modelandapprox} using the
Schrieffer-Wolf transformation which relates $J_s$, $J_t$ and $J_{st}$
to $V_a$ and $t_a$. For the symmetric Anderson model, $J_s=4 V^2_0/U$,
$J_{st}= 4t_aV_a/U$, and $J_t=4t^2_a/U$. Using the continuity
equation~(\ref{eq:continuityeq}), the current is
\begin{eqnarray}
\label{eq:kondocurrent}
I&=&\frac{2e}{\hbar} \;{\mathrm Im} \{ 
\sum_{kp\sigma} t_{kp} \langle c^\dagger_{k\sigma}c_{p\sigma}\rangle +
J_{st} \sum_{k p \sigma\sigma'} \langle c^\dagger_{k\sigma}
\vec s_{\sigma\sigma'}c_{p\sigma'} \cdot \vec S \rangle + \nonumber
\\ &+&
J_t \sum_{pp'\sigma\sigma'} \langle c^\dagger_{p\sigma}
\vec s_{\sigma\sigma'}c_{p'\sigma'} \cdot \vec S \rangle \}.
\end{eqnarray}
The first term is identical with the first term in
Eq.~(\ref{eq:currentfrequency}). In the lowest order of the tip-system
couplings ($t_{kp}, J_{st}, J_t$), the third term does not contribute.
The first term corresponds to the direct tip-substrate tunneling
channel -- process (1) in Fig.~\ref{fig:7} -- which includes the
scattering of conduction electrons from the local moment. The second
term corresponds to the direct tunneling into the magnetic impurity --
process (3) in Fig.~\ref{fig:7}. We note that the spin flip scattering
that gives rise to the Kondo effect is a higher order process. In the
lowest order, the channel (1) and the spin-flip component of (3) do
not give rise to interference because the final states have different
spin states. The lowest spin-flip process that does interfere with (1)
is of second order in $J$ and proportional to $J_s J_{st}$.

In the limit of large tip-metal separation, equivalent to the condition
$(J_s \gg J_{st} \gg J_t)$, the third term in
Eq.~(\ref{eq:kondocurrent}) as well as higher order contributions from
$J_{st}$ are neglected and all other exchange processes are
included in principle. This is equivalent to assuming that the 
state of the metal-adsorbate system is determined only by $J_s$
and is unaffected by the presence of the tip. Theoretically, the
problem then reduces to finding the spectral properties of the system
without the tip and using them in the expansion for tunneling via the
two terms in Eq.~(\ref{eq:kondocurrent}). 

As the system parameters move away from the Kondo limit -- that is
either $\epsilon_0$ shifts towards $\epsilon_{Fs}$ or $U$ becomes
smaller -- valence fluctuations appear. The Kondo resonance is then
due to both the spin and charge fluctuations. The separate energy
scale due to the spin fluctuations eventually disappears in the
mixed-valent regime and the Kondo peak merges with the broad resonance
centered at $\epsilon_0$. In the intermediate regime, where both
charge and spin fluctuations coexist on the impurity, another
tunneling channel exists. This channel is denoted by (2) in
Fig.~\ref{fig:7}. It also includes the contribution from higher order
non-flip processes similar to (3). We study the system in this regime
with the Hamiltonian defined in section~\ref{sec:modelandapprox}. 

We adopt the slave-boson technique of Coleman~\cite{Coleman84PRB} and
find the adsorbate Green's function using the non-crossing 
approximation (NCA)~\cite{Hewson93BOOK}. Following the theory of
section~\ref{sec:theorycurrent}, the final expression for 
current in Eq.~(\ref{eq:currentfinal}) is valid, as well as all the
consequent steps and approximation in~\ref{sec:theorycurrent}.
We insert the solution for the Green's function $G_a$ of the
$(U=\infty)$ interacting system in Eq.~(\ref{eq:conductancevsV}). This
is equivalent to including the three tunneling channels in
Fig.~\ref{fig:7} to lowest order in the tip-system couplings. We now
turn to the discussion of the results based on this approach.

\subsubsection{Results for large tip-substrate separation}
\label{subsubsec:Kondoequilibrium}

\begin{figure} 
\centerline{\epsfxsize=0.48\textwidth
\epsfbox{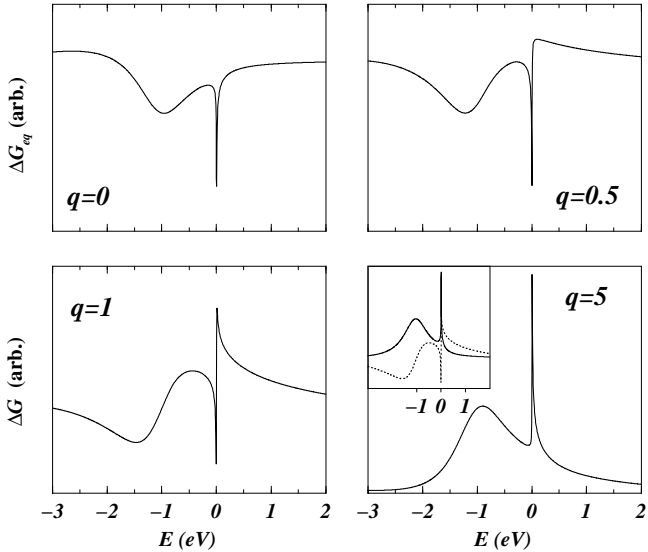}
}
\caption{The spectral lineshape for a model Kondo system described in
the text over large bias range that includes tunneling into the
broad resonance at $0.75$~eV below the Fermi level. Each panel
corresponds to different $q(\epsilon_{Fs})$ at the Fermi level. 
}
\label{fig:8}
\end{figure}

In order to model $Co/Au(111)$ studied both
experimentally~\cite{MadhavanetAl98S} and
theoretically~\cite{Kawasakaetal99JAP}, we choose a parameterization 
that gives the Kondo temperature $T_K \approx 70$~K appropriate for
the system. Our simplified model has degeneracy $N=2$ with no orbital
degeneracy, band width $2D=10$~eV, and the adsorbate level at
$\epsilon_0=0.75$~eV with the width $2\Gamma=1$~eV (the width of
a multiplet with an occupied level is $N\Gamma$ rather than
$\Gamma$!). We show the corresponding spectral function and the real
part of $G_a$ in the inset of Fig.~\ref{fig:8}.

Fig.~\ref{fig:8} shows the spectral properties of the system using
$\Delta {\mathcal G}_{eq}$  over the whole energy range of the
conduction band and 
for different values of $q$ at the Fermi level. The spectrum contains
information about the broad resonance at $\epsilon_0=0.75$~eV below
the Fermi level, as well as the prominent feature due to the Kondo resonance
at zero bias. We show the large bias voltage results only for
completeness since we do not expect the STM  experiments to be able to
provide spectroscopic information about the system over the whole
energy range shown.

The resonance lineshapes both in $Co/Au(111)$ and $Ce/Ag(111)$ correspond to
small values of $q$. Madhavan {\em et al.}~\cite{MadhavanetAl98S}
fitted the observed resonances to Fano lineshapes with $q\sim
0.7$. Our best fit would give approximately the same value of $q$.
In the case of $Ce/Ag(111)$, the observed feature is an almost symmetric
antiresonance corresponding to $q\sim 0$. Due to the contribution from the
substrate electronic structure to $q$, its value cannot be directly used
to make quantitative statements about the relative strength of the 
tunneling into the discrete state $d$ $(f)$ with respect to that
into the continuum. However, in agreement with
Li {\em et al.}~\cite{LietAl98PRL} and Lang~\cite{NDLang87PRL} we
conclude that the STM probes mostly the $sp$ wave functions and
the tunneling into the $f$-orbital is rather weak at the tip-adsorbate
distances used in the $Ce/Ag(111)$ experiment. The resonance is
mostly the result of interference between conduction electrons
scattering from the impurity. The larger value of $q$ in
$Co/Au(111)$ indicates stronger contribution from the coupling of the STM
to the $d$ orbital. This is expected because the $3d$ orbital is not as
tightly bound. 

The recent work of Kawasaka {\em et
al.}~\cite{Kawasakaetal98PLA,Kawasakaetal99JAP} deals with the spatial 
and spectroscopic profiles of the Kondo resonance. They begin with the 
Tersoff-Hamann~\cite{TersoffHamman85PRB} expression for
current~(\ref{eq:currentTersoffHamman}) and use
the local density of states given by~(\ref{eq:localdensity}). They
insert the self energy correction in the Green's function $G_a$ due to
the intra-adsorbate Coulomb correlations using perturbation theory
$(T > T_K)$ and Yamada's expansion in $U$ $(T < T_K)$ to study
the temperature dependence in the whole temperature range.
They neglect the additional temperature effects due to the Fermi
surface broadening, replace $(-\partial f/\partial \omega)$ by
the delta function, and evaluate the conductance at the tip bias.

One of the main conclusions of their work~\cite{Kawasakaetal99JAP} is
that the calculated temperature dependence of the resonance in the
differential conductance is indicative of the temperature dependence
of the Kondo resonance itself. They show results at the
experimentally relevant low temperatures for $Co/Au(111)$ and $Ce/Ag(111)$
in the range of temperatures $(T \leq 0.1 T_K)$. They find a rather
weak temperature dependence, due entirely to the temperature
dependence of the spectral function $\rho_a$. It is qualitatively the
same and comparable in magnitude with that found in Fig.~\ref{fig:6}
for a temperature independent resonance of the noninteracting system
for temperatures $T\leq \Gamma$. In our case, the temperature
dependence in $\Delta {\mathcal G}_{eq}$ is the consequence of the
Fermi surface broadening in the STM tip. Therefore a careful
deconvolution is necessary even at these low temperatures to extract
information about the temperature dependence of the Kondo
resonance. The other possibility is to eliminate variations in the
Fermi surface broadening of the tip.

\begin{figure} 
\centerline{\epsfxsize=0.48\textwidth
\epsfbox{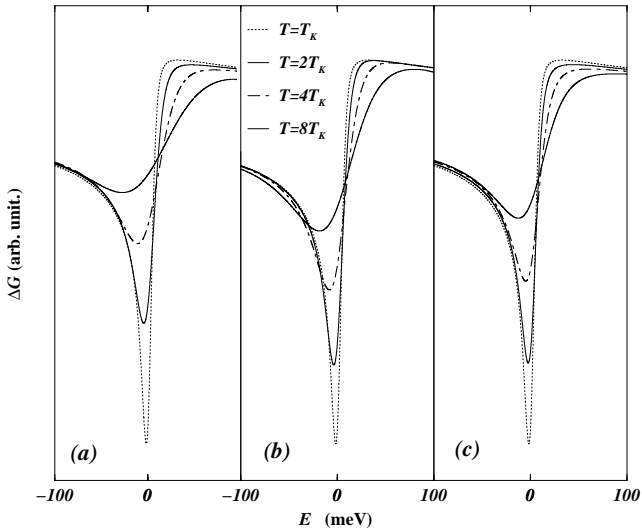}
}
\caption{Temperature dependence of the tunneling conductance through
the Kondo resonance. (a) the total dependence includes the temperature
dependence in the spectral function and the Fermi surface broadening; (b) 
temperature dependence due to the Fermi level broadening and with
spectral function given by its $T \approx T_K$ value for all
temperatures; (c) temperature dependence of only the spectral function
without the Fermi surface broadening (replaced by delta function). 
}
\label{fig:8temp}
\end{figure}

We show the temperature dependence for a Kondo system in
Fig.~\ref{fig:8temp}. Since the validity of our approximation is
limited to temperatures of order $T_K$ and higher, we show our results
only in this temperature range. Panel (c) shows the temperature
dependence one would 
observe with the tip at $T=0$~K and with varying substrate
temperature, i.e. when only the temperature dependence of the spectral
function is taken into account. Panel (b) assumes the substrate is at
a constant temperature $T=T_K$, which determines the shape of the
Kondo resonance, while the tip temperature is varied. We see that the
two contributions produce a very similar broadening of the Fano
resonance. Only a close look can uncover the difference. Panel (a)
shows the combined effect when the tip and substrate are kept at a
common temperature. Obviously, it would be difficult to determine the
contribution from the broadening of the Kondo resonance.

In addition to the temperature effects just discussed, Kawasaka {\em et 
al.}~\cite{Kawasakaetal98PLA,Kawasakaetal99JAP} also predicted the
existence of weak, long range oscillations in the current as a
function of the lateral tip position.  The particular long wavelength,
long range character of these predicted oscillations are a consequence
of their assumption that the tip-to-metal tunneling takes place into
the surface states of the (111) noble metal surfaces. The observed
resonances~\cite{MadhavanetAl98S,LietAl98PRL} are at variance with
these expectations.  On the other hand, the limited spatial extent of
the resonance observed at lateral tip positions up to $10$~\AA$\,$ is 
consistent with the rapid spatial decay determined by the bulk
$G_0^+$. No significant changes in the resonance lineshape are 
expected on this length scale (see discussion in 
section~\ref{subsec:surfacebandaverage} and Fig.~\ref{fig:5}). The
surface state would, however, be responsible for lineshapes variations
on larger length scales of order $20$~\AA. The fact that no resonance
is observed at such a distance from the impurity indicates that the
surface state contribution is indeed weak. Since Friedel
oscillations have been observed over long tip-impurity separations, we
believe that a weak tunneling resonance most likely persists in the
conductance over comparable distances, but more sensitive experiments
are necessary. In this case, changes in the lineshape with
$R_\parallel$ are expected. However, unlike Schiller and
Hershfield~\cite{SchillerHershfield00PRB}, we do not expect variations
in the lineshape due to the dominant contribution from the bulk states
on the length scale of $\le 5$~\AA$\,$ as discussed in
section~\ref{subsubsec:lateral}. 

\subsubsection{Nonequilibrium and hybridization effects at small
tip-substrate separation} 
\label{subsubsec:Kondononeqcurrent}

In typical STM experiments, the tip-substrate separation can be varied
from the point of contact where the tunneling resistance $R$ is a few
$100 k\Omega$ to distances where $R\sim 1 G\Omega$. Experimental
constraints limit the STM usefulness to the near Fermi level
spectroscopy -- especially at small $Z_t$ -- because of exponentially
increasing tunneling currents with bias. However, it is likely to be
possible to investigate the Kondo resonance -- which only requires
biases of the order of $\sim 10$~meV -- with very small tip-adsorbate
separations. It is therefore useful to analyze the physical
consequences of the small tip-metal separation on the resonance in
tunneling conductance.

In this case, nonequilibrium effects, as well as the tip-adsorbate
interaction, become important in the spectroscopy of Kondo
systems. First of all, as $\Gamma_{at}$ increases and becomes a
significant fraction of $\Gamma_{as}$ at small distances, the
tip-adsorbate hybridization will contribute to the width $\Gamma$ of
the resonance and to the renormalization of the level $\epsilon_0$. As
a result, the Kondo temperature, which depends sensitively on
$\Gamma$ and $\epsilon_0$, will change. This could be particularly
important for systems with very low bulk $T_K$, such as $Fe/Au$ with
$T_K\sim 1$~K. The Kondo temperature for an impurity adsorbed on the
surface of the metal is even lower than its bulk $T_K$ because the
lower coordination number for the adsorbate makes the width $\Gamma$
narrower. If $T_K \ll T$, the Kondo resonance will not be observed. In
certain systems and in the right temperature regime, it may be
possible for the Kondo resonance to reappear at smaller tip-adsorbate
distance as a result of the increased hybridization. This could also
be achieved by incorporating the adsorbate into the top surface
layer. The recent study of transition-metal impurities at the surface
of gold~\cite{Jamnealaetal00PRB} did not find any sign of the Kondo
effect in $V$, $Cr$, $Mn$, or $Fe$. We believe that, in the case of
iron, this is due to the low $T_K$ and may be an example of a
candidate system for the conditions discussed here.  On the other
hand, the smaller impurity-metal hybridization at the surface can lead
to magnetic behavior for systems which are nonmagnetic in the bulk,
such as $Ni/Cu$. There is a possibility for observing the transition
between magnetic and nonmagnetic behavior on a single system induced
either by embedding or by the proximity of the STM tip. 

\begin{figure} 
\centerline{\epsfxsize=0.48\textwidth
\epsfbox{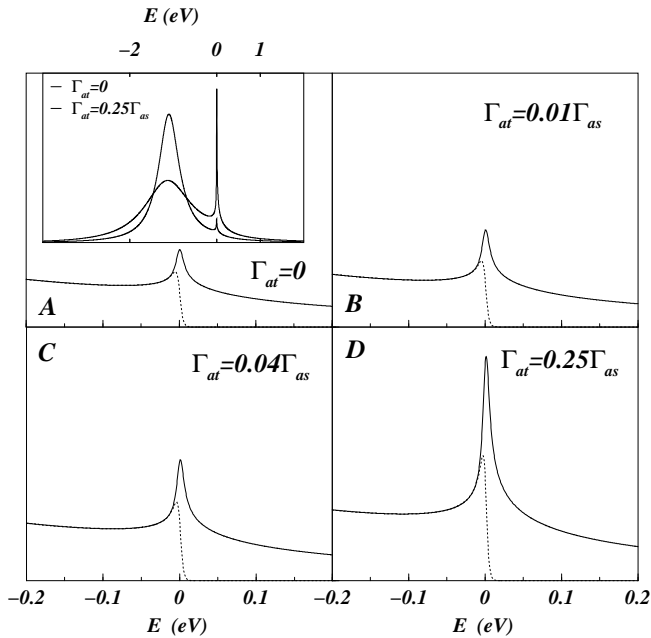}
}
\caption{The spectral function $\rho_a(\omega)$ at zero bias as a
function of the tip-substrate separation (defined in terms of
$\Gamma_{at}$). Dotted line= occupied density of states, bold line =
spectral density of states.
}
\label{fig:noneqKondo1}
\end{figure}

We show an example of the changing $T_K$ with hybridization in
Fig.~\ref{fig:noneqKondo1}, where the spectral function
$\rho_a(\omega)$ is plotted at zero bias as a function of the
partial width $\Gamma_{at}$, i.e. tip-metal separation for a model
system. We choose $D=5$~eV, $\Gamma_{as}=0.25$~eV, $\epsilon_a=-1$~eV,
and $T=30$~K. The Kondo temperature for this model in the limit
$t_a=0$, is $T_K\sim 30$~mK, much smaller than the temperature
$T$. Therefore the Kondo resonance in the spectral function is very
weak. When the tip is brought closer to the adsorbate, the Kondo
resonance acquires more spectral weight as the Kondo temperature
increases to $T_K\sim100$~mK at $\Gamma_{at}=0.01\Gamma_{as}$,
$T_K\sim0.2$~K at $\Gamma_{at}=0.04\Gamma_{as}$, and $T_K\sim1.5$~K at 
$\Gamma_{at}=0.25\Gamma_{as}$. Based on the justifications
in~\ref{app:adsorbate}, we neglected the effect of the direct 
metal-tip interaction on the spectral function and treat the effect of
the tip as another hybridization channel for the impurity
state.

The effect of varying hybridization -- due to either the presence of 
the STM tip or due to embedding or changing the environment of the
adsorbate -- on the tunneling resonance depends on the relation between
$T_K$ and $T$. For instance, when $T_K \ll T$, increased hybridization 
would produce stronger (sharper) tunneling resonance of the same width
since the spectral weight in the Kondo resonance increases while its
width remains almost constant until $T_K \sim T$. Also the experimental
resolution is limited by temperature in this regime. When, on the other
hand $T_K \ge T$, additional hybridization would not only increase the
spectral weight in the Kondo resonance but also its width. The two cases
should thus be distinguishable experimentally from each other and from the
possible lineshape variations with $Z_t$ as a result of changing $q$.

The second important effect of the strong tip-adsorbate interaction
is the breakdown of equilibrium relations at finite bias such
as the fluctuation-dissipation theorem
$G^<(\omega)=f(\omega)\rho_a(\omega)$ which consequently cannot be used
in deriving the expression for current~(\ref{eq:currentvsV}). This is
true in general because the electron occupation of the tip, metal, and
adsorbate electrons will no longer be thermal, i.e. will not be given
by $f_t(\omega)$ and  $f_s(\omega)$ but rather will be characterized
by a nonequilibrium distribution produced by the injected tunnel
electrons. The differential conductance is no longer proportional to
the local density of states and cannot be obtained using
Eq.~(\ref{eq:conductancevsV}). In Kondo systems, the hot electrons
not only modify the electronic distribution on the impurity, but also
modify the Kondo resonance itself. 

This is shown Fig.~\ref{fig:Kondospect} where the spectral function
and density of occupied states is plotted at selected bias
voltages in the limit of $|t_{pk}| \ll |t_{pa}|$.
The impurity has a resonance at $\epsilon_0=-1$~eV below the Fermi
level and total width $\Gamma=0.5$~eV produced by the hybridization
with both the tip and the substrate with the partial widths
$\Gamma_{at}=0.1\Gamma_{as}$. Temperature is of the order of $T_K$ in
this example. We see that the Kondo resonance broadens even more with
increasing bias. This is due to the increase in the rate of incoherent
scattering by $\sim eV_a/T$ -- an effect similar to temperature. At
the same time, the electron occupation develops a non-thermal profile
due to the large tip-adsorbate current. This is particularly visible
for negative biases where the density of states is larger.
Fig.~\ref{fig:Kondospect}(a) shows the equilibrium spectral function
(dotted) and the electron population on the resonance (solid bold). The
equilibrium spectral density is shown (dotted) in all panels. In
addition, the spectral density (solid) and occupation (bold solid) are
shown for the biases indicated in the figure by the labeled arrow. If
the coupling to the tip were comparable with the metal-adsorbate
hybridization, a double peak structure would develop. This has been
predicted by Wingreen and
Meir~\cite{WingreenMeir94PRB,SivanWingreen96PRB} in the context of the
nonequilibrium Kondo effect in quantum dots, also discussed by Plihal
et al.~\cite{Plihal00PRB1}. We see the onset of the double peak
structure in panels (b)-(d) where a small cusp develops at the
chemical potential of the tip. In summary, the bias has a significant
effect on the spectral density even when $\Gamma_{at} \sim 0.1
\Gamma_{as}$. 

\begin{figure} 
\centerline{\epsfxsize=0.48\textwidth
\epsfbox{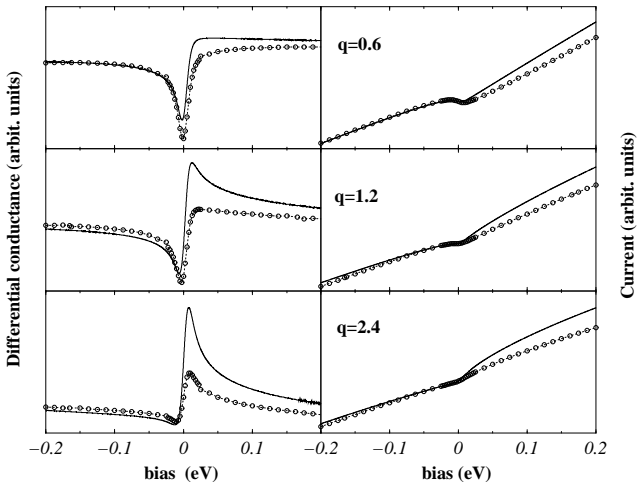}
}
\caption{Differential conductance (left) and current (right) for a
Kondo system defined in the text and with the spectral function
displayed in Fig.~\ref{fig:Kondospect}. Solid line is the
calculation in the lowest order in $t_{kp}$ and $t_{ap}$ and with the
equilibrium spectral function of Fig.~\ref{fig:Kondospect}. Circles
correspond to the nonequilibrium ${\mathcal G}_{tot}=dI_{tot}/dV$ and
$I_{tot}$ for $\Gamma_{at}=0.1\Gamma_{as}$. Each row corresponds to a
given value of $q$ at the Fermi level.
}
\label{fig:noneqKondo2}
\end{figure}

We show the tunneling current (right) and the corresponding
differential conductance (left) in Fig.~\ref{fig:noneqKondo2} for this
model of Kondo impurity and for the STM geometry defined by
$\Gamma_{at}=0.1\Gamma_{as}$ and $|t_{pk}| \ll |t_{pa}|$. The panels
correspond to $q=0.6$, $q=1.2$, and $q=2.4$, respectively. The current
on the right is calculated  using $I_{tot} = I_{eq} + \delta I_{non}$
of section~\ref{subsec:nonequilibrium} with $\delta I_{non}$ given by
Eq.~(\ref{eq:currentnoneq}). The differential conductance ${\mathcal
G}_{tot}$ on the left is obtained by differentiating the results
displayed on the right. It cannot be calculated from the expressions
in the text since the dependence of $\rho_a(\omega)$ and
$G_a^<(\omega)$ on the bias voltage modifies the contributions to the
current in a wide energy range, and ${\mathcal G}_{tot}$ is not
related in simple terms to the properties at the Fermi level of the 
tip. We compare $I_{tot}$ and ${\mathcal G}_{tot}$ (circles) with
$I_{eq}$ and ${\mathcal G}_{eq}$ (solid). The equilibrium quantities
were obtained in the lowest order in $t_{ap}$ and $t_{kp}$ and with
the equilibrium $G_a$ of Fig.~\ref{fig:Kondospect}(a).

We see that the broadening and disappearance of the Kondo resonance with
increasing bias at strong tip-adsorbate coupling is weakened in the
nonequilibrium calculation of ${\mathcal G}_{tot}$, because the contribution
$\delta I_{non}$ compensates partially for the spectral function effect.
The most consistent effect on the lineshape for various values of $q$
is the suppression of the resonance maximum and as a consequence a
more symmetric appearance. This behavior is
qualitatively different from both the hybridization effect and that of
the changing Fano parameter $q$ -- due to different decay constant of the 
impurity and metal states. Although the dependence of the tunneling resonance
on the tip-substrate separation $Z_t$ will contain all three contributions,
the hybridization and nonequilibrium contribution should only be important
at extremely small tip-adsorbate separations. The variations in $q$ should
not be important as it depends on the difference of the wavefunction tails
and the two remaining contributions should leave distinguishable signatures
in the tunneling resonance. It remains to be seen if the nonequilibrium 
condition play an important role in the tunneling between STM tip and Kondo
impurity.

Finally, we note that the limit $|t_{pk}| \ll |t_{pa}|$ discussed here
in connection with the nonequilibrium effects is not appropriate for
the recent STM experiments, where $t_{pa}$ is likely much weaker than
$t_{pk}$, even though the importance of $t_{pa}$ will increase
relative to $t_{pk}$ with decreasing $Z_t$. We will address the more 
general case in a future work.  

\section{Conclusions}
\label{sec:conclusions}

We used the Keldysh-Kadanoff method to study the spectroscopic
features of adsorbate resonances in the STM tunneling experiments. The 
central results of our theory are the general expression for current
Eq.~(\ref{eq:currentfinal}) to all orders in the tunneling matrix elements
and its equilibrium limit~(\ref{eq:currentfinalequilibrium}). Both are
valid for arbitrary intra-adsorbate electron correlations and thus apply to
both noninteracting ($U=0$), as well as magnetic (large $U$) systems.
The discussion of the Fano resonances is based on additional approximations
for the tunneling and hybridization matrix elements that lead to
expressions~(\ref{eq:currentvsV}) for the tunneling current and
Eq.~(\ref{eq:conductancevsV}) for differential conductance in 
the lowest order in the tip-to-system tunneling matrix elements $t_a$
and $t_c$, i.e. at large tip-surface separation, and the nonequilibrium
correction to the current in Eq.~(\ref{eq:currentnoneq}).

In the equilibrium limit, our theory of the tunneling current and
conductance differs from the standard theories of STM, in that the
dependence on LDOS is replaced by a tip specific quantity related
to the LDOS (Eq.~\ref{eq:Tt}). The current is expressed
entirely in terms of the adsorbate Green's function $G_a^R$, the tip
density of states, the tunneling matrix elements and the substrate
Green's function. We used the formulation to study the resonance lineshape
as a function of temperature, tip-substrate separation, and lateral
tip position. We summarize our findings as follows. 

(1) The role of impurity state resonances in tunneling can be discussed
in terms of two limiting cases. When direct tunneling across the
barrier is weak, the resonance within the barrier provides an
additional tunneling channel and can significantly enhance the
tunneling current. This is the case of quantum dots in Coulomb
blockade regime. If on the other hand, the tunneling into the
continuum is strong, the presence of an ``impurity'' state could 
suppress the tunneling current due to the additional scattering of the
conduction electrons in the metal from the impurity, i.e. increased
resistance. The tunneling into the Kondo resonance in the recent STM
experiments seems to be closer to the latter limit.

(2) The information about electron correlations
and the Kondo resonance enters the tunneling problem through the
impurity Green's function $G_a$ while the position dependence of 
the conductance is controlled by the electronic structure of
the metal.

(3) The spatial decay of the observed Fano resonance in the
recent experiments~\cite{MadhavanetAl98S,LietAl98PRL} is consistent
with the conclusion that tunneling into the bulk conduction and
hybridized $sp$ impurity states gives rise to most of the signal.
The absence of any observable resonance at distances larger than
$\sim 10$~\AA$\,$ suggests that the contribution from the surface state
on Au(111) and Ag(111) to the resonant tunneling is not important in
these experiments. However, the surface states are important in
special cases, as indicated by the recent corral
experiments~\cite{Manoharan00Nat} in which the contribution of 
the surface states is enhanced by scattering from the walls of the 
corral.

(4) At large $Z_t$, tunneling into conduction states with
$k_\parallel$ having the smallest parallel component corresponding to
energy $\omega=\epsilon_{k_\perp}+\epsilon_{k_\parallel}$ is
strongly favored. This leads to the disappearance of the
current oscillation vs. the lateral tip position due to tunneling into
the bulk states which should otherwise be observed with period of
about $1-2$~\AA$\,$ (corresponding to the bulk $k_F$) for typical
experimental conditions. Therefore no oscillations in the lineshape
should be observed on this length scale for typical tip-surface
separation. The occurrence of an antiresonance with tip position at
certain neighboring sites predicted by Schiller and
Hershfield~\cite{SchillerHershfield00PRB} has its origin in these 
oscillations. It is a result of a simplified model for the surface
electronic structure and we believe is unphysical. The small current
oscillations predicted by Kawasaka {\em et
al.}~\cite{Kawasakaetal99JAP} assume that the surface states are
all-important in the spatial dependence of the resonance which seems
to contradict the experimental results. We believe the surface state
should be important at larger distances since on the (111) noble metal
surfaces $k_F \sim 0.15 - 0.2$~\AA$^{-1}$ and the corresponding period
of oscillations is about $20$~\AA$\,$ (as observed experimentally as
Friedel oscillations). We expect changes in the resonance
lineshapes with this spatial period if the contribution from the 
surface state if the signature of the resonance is detectable at such
distances.

(5) From the lineshapes observed in $Co/Au(111)$ and $Ce/Ag(111)$, we
conclude that the direct tunneling into the discrete ($d$ or $f$)
state is quite weak -- stronger in $Co/ Au(111)$. This confirms that the
STM is mostly a probe of the delocalized $sp$ states and couples only
weakly to the tightly bound $d$ or $f$ orbitals at typical tip-surface
separations. Therefore the dominant process giving rise to the resonance
lineshape is the tip-to-metal tunneling and interference between
conduction electrons scattering from the local moment.  

(6) The temperature dependence in differential conductance does not
reflect only the temperature dependence of the Kondo resonance, but
includes also the effect of Fermi surface broadening (mostly of the tip).
The two contributions are of the same order of magnitude and 
qualitatively indistinguishable. Therefore, the temperature dependence
in the differential conductance cannot be used directly to make
conclusions about the temperature dependence of the resonance without
controlling the tip Fermi surface broadening or without deconvolution.

(7) At small tip-surface separations, nonequilibrium effects as well as
the additional tip-adsorbate hybridization may play an important role
-- especially in Kondo systems. The main effect of the finite bias
voltage in this case is to broaden the Kondo resonance and produce
nonequilibrium electron population on the adsorbate. The observed 
Fano resonance in differential conductance also broadens and its
maximum is suppressed. The effect of the tunneling current on the
Kondo resonance should thus leave a characteristic dependence of the
lineshape on $Z_t$. 

\section{Acknowledgments}

We thank R. Celotta, E. Hudson, M. Stiles, and J. Stroscio for fruitful
discussions and for helping us understand the experimental issues more
clearly.

\setcounter{section}{0}
\appendix{Adsorbate Green's function}
\label{app:adsorbate}

An important quantity in the theory of tunneling current through
adsorbate resonances is the adsorbate Green's function $G_a$. Using
the equation of motion method, we find the expression for $G_a$
defined as the Fourier transform of  
\begin{equation}
\label{eq:adsorbgreen}
G_a(t,t')=-i \langle T_C c_a(t) c_a^\dagger(t') \rangle.
\end{equation}
We do this for the case of arbitrarily strong coupling between the tip
and the adsorbate with the intent to describe the nonequilibrium
effects at finite bias. However, in this paper we consider the effect of
the direct tip-metal interaction on $G_a$ to be weak and neglect
it. Extension to the full description will be considered in future
work. We believe the approximations adopted here capture the most
important nonequilibrium effects. 

We discuss both the noninteracting $(U=0)$ and interacting
$(U=\infty)$ model. Since the solution in both limits for the
adsorbate-metal interaction is well known, we limit our discussion to
the issues specific to the addition of the biased tip and refer reader
to standard texts for the details. The $(U=\infty)$ model is solved
using the slave boson technique and NCA. In this approach a new
pseudofermion is introduced by the transformation $c_a \rightarrow c_a
b^\dagger$ in the Hamiltonian~(\ref{eq:SystemHamiltonian}), where
$b^\dagger$ is the creation operator for the slave boson. This
eliminates the interaction term $U$ from the Hamiltonian as discussed
by Coleman.~\cite{Coleman84PRB}

The time ordering operator $T_C$ orders the time according to their
position on contour in the complex time plane~\cite{LangrethBOOK}. It
is important to note that the equations must be first solved in the
complex time domain and then analytically continued to the real axis
as was pointed out in the previous appendix. The analytic continuation
is performed before the Fourier transform, so we must be careful about
how we deal with the Fourier transformed equations. Relevant details
are in~\ref{app:analytic}. Here we discuss the equations of motion 
satisfied by the Fourier transforms of the time ordered Green's
functions and the analytic continuation is performed at the end
according to the rules in~\ref{app:analytic}. All Green's functions
and self-energies in the following expressions are function of
frequency $\omega$ and, therefore, we omit their argument to simplify
the notation.
\begin{figure} 
\centerline{\epsfxsize=0.48\textwidth
\epsfbox{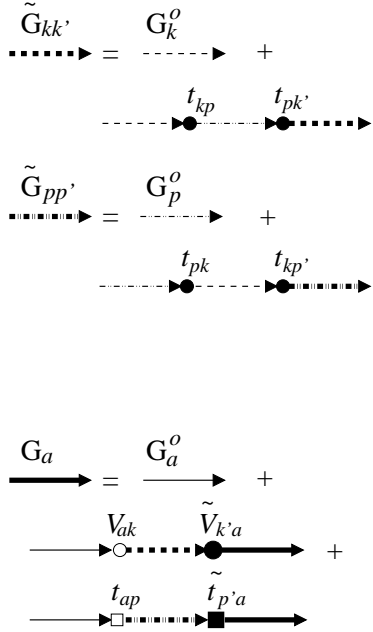}
}
\caption{(a) Diagrammatic expansion for the metal ($\tilde G_{kk'}$)
and tip ($\tilde G_{pp'}$) Green's functions in mutual interaction but 
without the adsorbate. (b) diagrammatic expansion for the adsorbate
Green's function $G_a$ of the noninteracting Anderson model using the
solutions $\tilde G_{kk'}$ and $\tilde G_{pp'}$.
}
\label{fig:diagram1}
\end{figure}

The Green's function for the impurity state $G_a$ can be written in a
standard way 
\begin{equation}
\label{eq:greend}
G_a = (\omega-\epsilon_0-\Sigma_a)^{-1},
\end{equation}
using the self energy $\Sigma_a(\vec R_t, Z_0;\omega)$. The solution
for $G_a$ is thus reduced to finding $\Sigma_a$. We first treat a
closed shell or nonmagnetic open shell ($V_{ka} \gg U$) adsorbate for
which electron correlations can be neglected. We begin by considering
the tip-substrate system without the adsorbate. We define $\tilde
G_{kk'}$ and $\tilde G_{pp'}$ in analogy with
$G_a$~(\ref{eq:adsorbgreen}) as the Green's functions of the metal and
tip states, respectively, in the absence of the adsorbate. These are
not identical with the Green's functions $G_{kk'}$ and $G_{pp'}$ for
the full system introduced in~\ref{app:equationofmotion}. The bare
metal-tip system is described by the Hamiltonian of
section~\ref{sec:modelandapprox} with
$\epsilon_0=U=t_{ap}=V_{ak}\equiv0$. Using the equations of motion, we
can write 
\begin{equation}
\label{eq:gkk'}
(\omega-\epsilon_k) \, \tilde G_{kk'}
=\delta_{kk'} + \sum_{k''}\Sigma_{kk''}\tilde G_{k''k'}
\end{equation}
and
\begin{equation}
\label{eq:gpp'}
(\omega-\epsilon_p) \, \tilde G_{pp'}= \delta_{pp'}  +
\sum_{p''}\Sigma_{pp''}\tilde G_{p''p'}
\end{equation}
where the self-energies are $\Sigma_{kk'}=\sum_p t_{kp}G^0_pt_{pk'}$
and $\Sigma_{pp'}=\sum_k t_{pk}G^0_kt_{kp'}$ and
$G_k^0=(\omega-\epsilon_k+i\eta_k)^{-1}$ and
$G_p^0=(\omega-\epsilon_p+i\eta_p)^{-1}$ are the Green's functions for
the clean metal and tip, respectively, without their mutual
interaction. The coupled equations are shown diagrammatically in
Fig.~\ref{fig:diagram1}. The solutions for $\tilde G_{kk'}$ and
$\tilde G_{pp'}$ can be formally written as the inverse of
$D_{kk'}=\delta_{kk'}(\omega-\epsilon_k)-\Sigma_{kk'}$ and
$D_{pp'}=\delta_{pp'}(\omega-\epsilon_p)-\Sigma_{pp'}$.
We also define the adsorbate-metal and adsorbate-tip
hybridization matrices modified by the tip-substrate interaction as
$\tilde V_{ka} = V_{ka} + \sum_p t_{kp} G_p^0 t_{pa}$
and $\tilde t_{pa} = t_{pa} + \sum_k t_{pk} G_k^0 V_{ka}$.
With these definitions and with $\tilde G_{kk'}$ and $\tilde G_{pp'}$
obtained through~(\ref{eq:gkk'}) and~(\ref{eq:gpp'}), the solution for the 
noninteracting $\Sigma_a$ -- shown diagrammatically in Fig.~\ref{fig:diagram1} --
is formally given by
\begin{equation}
\label{eq:selfenergy}
\Sigma_a= \sum_{kk'}V_{ak}\tilde G_{kk'}\tilde V_{k'a} +
\sum_{pp'}t_{ap} \tilde G_{pp'} \tilde t_{p'a}.
\end{equation}
The evaluation of the self energy $\Sigma_a$ is rather complicated in the
general case of strong tip-to-substrate coupling. We proceed with formulation
of the general nonequilibrium theory for the tunneling current using this self
energy (section~\ref{subsec:generaltunneling}) and then we discuss two
limiting cases: (a) the equilibrium limit $|t_{kp}|, |t_{ap}| \ll |V_{ka}|$
(section~\ref{subsec:tunnelcurrent}) in which case the second term
in~(\ref{eq:selfenergy}) is neglected and $\tilde V_{ka}$, $\tilde G_{kk'}$
replaced by $V_{ka}$, $G_k^0\delta_{kk'}$;
and (b) the nonequilibrium case under the assumption
$|t_{kp}| \ll |t_{ap}| \sim |V_{ka}|$, in which case we keep both terms
in~(\ref{eq:selfenergy}) and replace $\tilde V_{ka}$, $\tilde t_{ap}$,
$\tilde G_{kk'}$, and $\tilde G_{pp'}$ by
$V_{ka}$, $t_{ap}$, $G_k^0 \delta_{kk'}$, and $G_p^0 \delta_{pp'}$.
Section~\ref{subsubsec:Kondononeqcurrent} deals with tunneling through a Kondo
impurity in this limit. The case (b) includes the effect of the increased
hybridization of the discrete state due to the tip presence and the onset
of nonequilibrium population on the adsorbate at finite bias. 

In order to study these corrections in the limit (b), $(|t_{pk}| \ll
|t_{pa}| \sim |V_{ka}|)$, we replace the Green's functions
$\tilde G_{kk'}$ and $\tilde G_{pp'}$ by the noninteracting ones, i.e.
$\tilde G_{kk'}=\delta_{kk'}G^0_k$ and $\tilde G_{pp'}=\delta_{pp'}G^0_p$
and the modified $\tilde V_{ka}$ and $\tilde t_{pa}$
by $V_{ka}$ and $t_{pa}$. The self energy $\Sigma_a$ then simplifies to
\begin{equation}
\label{eq:selfenergyapprox}
\Sigma_a^0= \sum_k |V_{ak}|^2 G_k^0 + \sum_p |t_{ap}|^2 G_p^0.
\end{equation}
The largest source of error in writing the approximate self energy
is the neglect of the possibly significant interference effects at larger
$t_{kp}$ as a result of the phase difference between $\tilde t_{pa}$ and $t_{pa}$.
It is always reasonable to replace $\tilde V_{ka}$ by $V_{ka}$, as long as the
adsorbate is on the surface rather than on the STM tip. These general case
will be the topic of a future study. If the tip distance from the adsorbate
is much larger than the adsorbate-metal separation, so that $t_{ap}
\ll V_{ak}$, the self-energy is well described by the first term
only. In such a case, the STM does not strongly modify the studied system.
It is then reasonable to characterize the system without the presence of the
STM tip and then consider the tunneling.

\begin{figure} 
\centerline{\epsfxsize=0.48\textwidth
\epsfbox{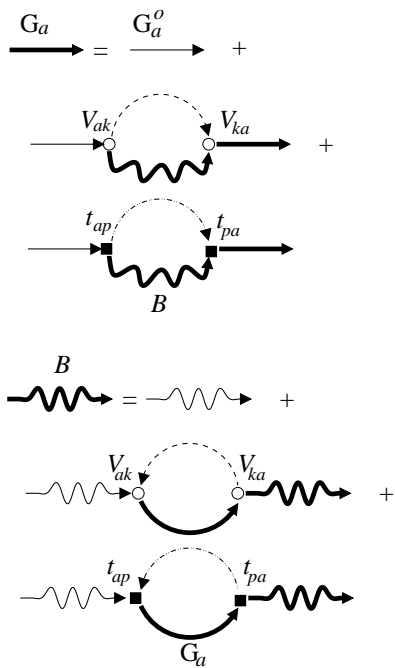}
}
\caption{The diagrammatic expansion for the adsorbate Green's
function $G_a$ of the interacting Anderson Hamiltonian ($U=\infty$)
within the NCA.
}
\label{fig:diagram2}
\end{figure}

Finally, we discuss the Green's function $G_a$ in the limit (b) for a Kondo
impurity which is likely to show stronger dependence on the bias and tip
interaction. We find the nonequilibrium Green's function $G_a$ under the
same assumption that lead
to $\Sigma_a^0$ for the noninteracting Anderson Hamiltonian. We solve the
interacting system in the limit of $U=\infty$ using the NCA approximation,
shown diagrammatically in Fig.~\ref{fig:diagram2}. The self
energy is not a simple sum of the two contributions from the metal and tip
as it was in the noninteracting system, because the occupation of the resonance
is limited to one electron and the hybridization is now correlated -- formally
through the slave boson Green's function $B(\omega)$. The two coupled equations
in Fig.~\ref{fig:diagram2} are solved selfconsistently. The expressions for
the NCA self energy is obtained in a standard with the help of the diagrams
in Fig.~\ref{fig:diagram2}.~\cite{Hewson93BOOK}

\appendix{Equations of motion for $G_{pk}$ and $G_{pa}$}
\label{app:equationofmotion}

In this appendix, we find the solution for $(\sum_{kp} t_{kp}
G_{pk}+\sum_{ap}t_{ap}G_{pa})$ entering the expression for tunneling
current~(\ref{eq:currentfrequency}) for the general case of arbitrary
tip-system coupling. Ultimately, the interesting regime in connection with
typical STM experiments is one in which $t_{kp}, t_{ap} < V_{ak}$. However,
we want to be able, in principle, to study the system when the coupling of the
STM tip to the system and the tunneling current are strong. This
creates nonequilibrium occupation on the adsorbate resonance and modifies the
spectroscopic properties of the system. We therefore proceed by deriving the
most general expression valid for arbitrary coupling strength $t_{kp}$ and
$t_{ap}$ and discuss an approximation (b) that allows us to take into
account the most important nonequilibrium effects as described in the previous
appendix. For this purpose we introduce Green's function  
$G_{pa}(\omega)$, $G_{pk}(\omega)$, $G_{kk'}(\omega)$,
$G_{pp'}(\omega)$, and $G_{ka}(\omega)$ as the Fourier transform of 
\begin{equation}
\label{eq:greenapk}
G_{ij}(t,t')=-i \langle T_C c_i(t) c_j^\dagger(t') \rangle.
\end{equation}

We now turn to the equations of motion for the Green's functions
relevant for the tunneling current. The following
expressions are valid for arbitrary interaction $U\neq 0$ and the
nature of the intra-adsorbate interactions are contained fully in the
solution for $G_a$ discussed in the previous appendix. The 
first term in the current~(\ref{eq:currentfrequency}) contains the
tip-adsorbate propagator which satisfies 
\begin{equation}
\label{eq:greenpd}
(\omega-\epsilon_p) \, G_{pa}= t_{pa}G_a+\sum_k t_{pk}G_{ka}.
\end{equation}
It is expressed in term of $G_a$ already solved within a given
approximation in the previous appendix through~(\ref{eq:greend}) and
in terms of the metal-adsorbate Green's function
\begin{equation}
\label{eq:greenkd}
(\omega-\epsilon_k) \, G_{ka}= V_{ka} G_a+\sum_p t_{kp}G_{pa}.
\end{equation}
The last two equations are coupled and need to be solved
self-consistently. We do this by substituting Eq.~(\ref{eq:greenkd})
for $G_{ka}$ in~(\ref{eq:greenpd}), and vice versa. The solutions are 
then expressed in terms of $G_a$, $\tilde G_{pp'}$, and $\tilde t_{pa}$
discussed in the previous appendix as
\begin{equation}
\label{eq:greenpdsimplified}
G_{pa}= \sum_{p'} \tilde G_{pp'} \tilde t_{p'a} G_a.
\end{equation}
We will also need the solution for $G_{ak}$. The
tip-induced correction to $V_{ka}$ contributes to the phase of
$V_{ka}$, as well as its magnitude, and could thus affect the
lineshape significantly in the strong coupling limit. But it should be
particularly weak when $t_{kp}, t_{pa} \ll V_{ka}$ and it will be safe
to ignore it. We write 
\begin{equation}
\label{eq:greenkdsimplified}
G_{ak}= G_a\sum_{k'} \tilde V_{ak'} \tilde G_{k'k}.
\end{equation}
The second term in~(\ref{eq:greenkd}) is negligible when the
tip-adsorbate separation is much larger than the adsorbate-metal
separation. Neglecting this term is equivalent to replacing $\tilde
G_{pp'}\rightarrow G_p^0\delta_{pp'}$ in~(\ref{eq:greenpdsimplified})
and $\tilde G_{kk'}\rightarrow G_k^0\delta_{kk'}$, $\tilde
V_{ka}\rightarrow V_{ka}$ in~(\ref{eq:greenkdsimplified}). The
tip-adsorbate Green's function $G_{pa}$ is then expressed entirely in 
terms of $G_a$ and the unperturbed conduction electron Green's functions.

The second term in~(\ref{eq:currentfrequency}) contains the tip-metal
propagator $G_{pk}$, which satisfies 
\begin{equation}
\label{eq:greenpk}
(\omega-\epsilon_p) \, G_{pk}=\sum_a t_{pa}G_{ak}+\sum_{k'}t_{p k'}G_{k'k}.
\end{equation}
It is expressed in terms of $G_{ak}$~(\ref{eq:greenkdsimplified}) discussed in 
the previous paragraph and in terms of $G_{k'k}$, the Green's function
for the substrate conduction electrons
\begin{equation}
\label{eq:greenkk'}
(\omega-\epsilon_{k'}) \, G_{k'k}=\delta_{kk'} + 
\sum_a V_{k'a} G_{ak}+\sum_p t_{k'p}G_{pk}.
\end{equation}
We see that $G_{kk'}$ couples to $G_{pk}$,~(\ref{eq:greenpk}), and
also to $G_{ak}$,~(\ref{eq:greenkdsimplified}), already solved in
terms of $G_a$ and $\tilde G_{kk'}$. The last two equations can be
solved self-consistently to give 
\begin{equation}
\label{eq:greenkk'simplified}
G_{k'k}=\tilde G_{k'k} + \sum_{k_1k_2} \tilde G_{k'k_1} \tilde V_{ak_1}
G_a \tilde V_{k_2a} \tilde G_{k_2k}.
\end{equation}
and 
\begin{equation}
\label{eq:greenpksimplified}
G_{pk}= \sum_{p'} \tilde  G_{pp'} \, ( t_{p'k} + \sum_{ak'}
\tilde t_{p'a} G_a \tilde V_{ak_1} )\,  G_k^0.
\end{equation}
For the purpose of analytic continuation, it is important to keep track
of the order in which the Green's
functions appear in the product in the above equations. The ``lesser''
Green's functions are then obtained according to rules stated
in~\ref{app:analytic}. The equilibrium limit of the theory is achieved
by neglecting the last term in~(\ref{eq:greenkk'}) along with the
equivalent approximations for $G_a$ and $G_{ak}$ discussed above. This
removes the self-consistency requirement and neglects the effect of
the tip on the substrate conduction electrons, but not on the
tunneling current. The solution for $G_{kk'}$ is then identical to
that of the system without the tip.

\appendix{Rules for dealing with nonequilibrium Green's functions in
frequency space} 
\label{app:analytic}

In this appendix we review the process of analytic continuation of the
complex time contour expression to integrals on the real time axis. We
follow Langreth's generalization~\cite{LangrethBOOK} of
Kadanoff-Baym's method~\cite{KadBaym} described in detail by Haug and
Jauho~\cite{JauhoBOOK}. Four Green's functions appear in the
nonequilibrium theory ``lesser'' $G^<$, ``greater'' $G^>$, retarded
$G^R$, and advanced
$G^A$~\cite{KadBaym,LangrethBOOK,JauhoBOOK,MahanBOOK}. We frequently
need to find the retarded (advanced) and ``lesser'' Green's function
corresponding to Green's function $A$ time ordered in the complex time
plane expressed as a product of $N$ time ordered functions
$B...Z$ , i.e. 
\begin{equation}
\label{eq:greencomplex}
A(t,t')=\int_C d\tau_1 ... d\tau_2 B(t,\tau_1).. Z(\tau_2,t') 
\end{equation}
where all functions are assumed fermion-like. 
The desired expressions analytically continued onto the real time axis
are~\cite{LangrethBOOK} 
\begin{equation}
\label{eq:greenanalyticret}
A^{R(A)}(t,t')=\int_{-\infty}^\infty d\tau_1 ... d\tau_2
B^{R(A)}(t,\tau_1) ... Z^{R(A)}(\tau_2,t') 
\end{equation}
which consists of only one term and
\begin{eqnarray}
\label{eq:greenanalyticless}
&&\glsup{A}(t,t')=\int_{-\infty}^\infty d\tau_1 ... d\tau_3
... d\tau_2 \\ \nonumber
&& [ ... + B^R(t,\tau_1)... \glsup{C}(t,\tau_3)...Z^A(\tau_2,t')
+ ... ]
\end{eqnarray}
where each of the $N$ terms in the integral has exactly one function
of the type $\glsup{f}$, all functions to the left (right) of it are
retarded (advanced), and each of the terms has the $\glsup{f}$ in a
different position.

We are dealing here with the case of time independent
perturbations (steady state current). The double time propagators
then only depend on the time difference ($t-t'$). The
equations~(\ref{eq:greenanalyticret}),~(\ref{eq:greenanalyticless}) 
can be Fourier transformed and we can write the time ordered
Green's function $A(\omega)$ as a simple product
\begin{equation}
\label{eq:greenfrequency}
A(\omega)=B(\omega)...Z(\omega).
\end{equation}
The rules for writing the expression for the ``lesser'' and retarded
function in the frequency space are then directly carried out by 
Fourier transforming the equations~(\ref{eq:greenanalyticret})
and~(\ref{eq:greenanalyticless}). Leaving out the frequency arguments, 
we write
\begin{equation}
\label{eq:greenfrequencyanalyticret}
A^{R(A)}= B^{R(A)}...Z^{R(A)}
\end{equation}
and 
\begin{equation}
\label{eq:greenfrequencyanalyticless}
\glsup{A}= ... + B^R...\glsup{C}...Z^A
+ ... 
\end{equation}
We note that it is important to keep track of the order in which the
function $B ... Z$ appear in the time integral when the Fourier
product is formed.

\end{document}